\documentclass[useAMS,usenatbib,usegraphicx]{mn2e}

\bibpunct{(}{)}{;}{a}{}{,}

\newcommand{\ignore}[1]{}
\newcommand{\eq}[1]{\begin{equation} #1 \end{equation}}
\newcommand{\eql}[2]{\begin{equation} \label{#1} #2 \end{equation}}

\newcommand{\eqs}[1]{\begin{eqnarray} #1 \end{eqnarray}}

\newcommand{\change}[1]{#1}
\newcommand{\rvtwo}[1]{#1}

\usepackage{mathptmx}
\usepackage{amssymb}
\title[MHD stability of BLR clouds]{Magnetohydrodynamic stability of
  broad line region clouds}
\author[M. Krause et al.]{Martin Krause$^{1,2}$\thanks{E-mail:
krause@mpe.mpg.de,mkrause@usm.lmu.de}, Marc Schartmann$^{1,2}$, 
Andreas Burkert$^{1,2,3}$\\
$^1$Universit\"atssternwarte M\"unchen, Scheinerstr.~1, 81679 M\"unchen, Germany\\
$^2$Max-Planck-Institut f\"ur Extraterrestrische Physik, Postfach 1312, Giessenbachstr., 85741 Garching, Germany \\
$^3$Max-Planck-Fellow\\}
\begin{document}

\date{Accepted \date. Received 2011 December 19}

\pagerange{\pageref{firstpage}--\pageref{lastpage}} \pubyear{2008}

\maketitle

\label{firstpage}

\begin{abstract}
\change{
Hydrodynamic stability has been a longstanding issue for the cloud
model of the broad line region in active galactic nuclei. 
We argue that the clouds \rvtwo{may be}
gravitationally bound to the supermassive black hole. If true,
stabilisation by thermal pressure alone becomes even more
difficult. We further argue that if
magnetic fields should be present in such clouds at a
level that could affect the stability properties,
they need to be strong enough to compete with the radiation pressure
on the cloud. This would imply magnetic field values of \rvtwo{a few}
Gauss for a sample of Active Galactic Nuclei we draw from the literature.
}

\change{We then investigate the effect of several magnetic
  configurations on cloud stability} 
in axisymmetric magnetohydrodynamic simulations.
For a purely azimuthal magnetic field
which provides the dominant pressure support, \change{the
cloud}
first gets compressed by the opposing radiative and gravitational
forces. The pressure inside the cloud then increases, and it expands
vertically. 
Kelvin-Helmholtz and column density instability
%and repeated compression and expansion events 
lead to a
filamentary fragmentation of the cloud. This radiative dispersion
continues until the cloud is shredded down to the resolution
level. 
%Mixing with the ambient pressure supported gas then leads to a
%loss of rotational support and the cold filamentary pieces of the
%original cloud fall towards the centre.
For a helical magnetic field configuration, a much more stable cloud core survives with a
stationary density histogram which takes the form of a power law.
Our simulated clouds develop sub-Alfv\'enic internal motions on the
level of a few hundred~km/s. 
\end{abstract}

\section{Introduction}\label{intro}
Broad emission lines are produced in the immediate vicinity of
optically active super-massive black holes \citep[SMBH, for reviews see
e.g.][]{Ost88,Pet97,Net08}. They may be used to infer the black hole
mass in galaxies with such an Active Galactic Nucleus \citep[AGN,
e.g.][]{Benea09}, and are a standard part of optically active AGN.
The basic line emitting entity is usually referred to as a cloud. 
The emission mechanism is photoionisation
by the central parts of the accretion disc.  Photoionisation models
\change{predict} the clouds to have a typical temperature of order
$10^4$~K, number densities of $n_\mathrm{cl}=10^{10\pm1}$~cm$^{-3}$, 
sizes of $R_\mathrm{cl} = 10^{12\pm1}$~cm and column
densities of \change{$N_\mathrm{cl}>2\times 10^{22}$~cm$^{-2}$}
\citep[e.g.][]{KK81,FE84,RNF89}.
\change{Further important constraints come from
  reverberation mapping \citep[e.g.][see below for a comparison of
  results of these two methods]{Pet88}.}
The complete physics of these clouds is however highly complex and involves
also pressure, radiative, centrifugal, gravitational, and magnetic
forces, probably on a very similar level. We are not aware of any
attempt to include all these processes into a single model, but
different authors have looked at some particular aspects of the
problem:
A general assumption for the cloud ensemble is often virial
equilibrium. Since the gravitational potential is dominated by the
SMBH, this would imply Kepler orbits. This treatment neglects the
contribution of radiation
pressure to the dynamics. The latter restriction has been relaxed in a
recent series of papers (\citeauthor{Marcea08} \citeyear{Marcea08},
\citeyear{Marcea09}; \citeauthor{Net09}
\citeyear{Net09}, \citeyear{Net10}; \citeauthor*{KBS11}
\citeyear{KBS11} (hereafter: paper~I)). The general finding is that the radiation force may
contribute significantly, and that in this case, the clouds on bound
orbits could be significantly sub-Keplerian. For low angular momentum
(strong radiation pressure support), the orbits have to be highly
excentric.
 
\change{The importance of radiation pressure has actually been
  recognised early on \citep[e.g.][]{TMcK73,McKT75,BM75,BM79,Mat76}.
In fact, many of the earlier articles study the possibility that
radiation pressure is dominant, also compared to gravity, 
and the clouds are unbound. In this case, it would however not be easy
to understand why the black hole masses derived from reverberation
mapping under the assumption of virial equilibrium may be brought into
agreement with the correlation between black hole mass and host galaxy
properties with a uniform scaling factor \citep{Onkea04}. 
For optically thin clouds, one would
expect almost no response to the variability of the ionising
continuum. The strong variability of the emission lines, lead by
variations in the continuum is therefore evidence for the presence of
optically thick clouds \citep[e.g.][]{SG07}. However, the relative importance
of gravity versus radiation pressure is proportional to the optical
thickness (compare equation (\ref{prg_ratio}), below).\newline This opened up
the possibility
 that the Broad Line Region (BLR) is gravitationally bound and possibly
disc-like with a significant total angular 
momentum. In fact, several pieces of evidence that point in this direction have been found
over recent years}
(compare paper~I and references therein). One recent piece of
evidence comes from spectropolarimetry, where the BLR is spatially
resolved by an equatorial scattering region, leading to different
polarisation angles in the red and blue wings of
emission lines \citep{Smea05}. Most recently, \citet{KolZet11} have
shown that the shape of the broad emission lines in many objects 
may be well fit with
the assumption of a turbulent thick disc.

Cloud stability and confinement is a long-standing issue \citep[for a review]{OM86}:
The clouds should be in rough pressure equilibrium with their
environment 
(e.g. \citeauthor*{KMKT81} \citeyear{KMKT81}; \citeauthor{Krolik88} \citeyear{Krolik88}).
%\citep[e.g.][]{KMKT81,Krolik88}. 
To reach the required pressure of $p\approx10^{-2}$~dyne~cm$^{-2}$, 
the inter-cloud medium needs either a high temperature
\citep[$>4\times10^7$~K, ][]{Krolik88}   
 or a strong magnetic field \citep{Rees87}.
Apart from the confinement issue, the clouds should be
hydrodynamically unstable:
\citet{Mat82} has shown that while optically thin clouds may be close to uniformly
accelerated by the radiation pressure, there remain internal radial
pressure imbalances, which especially in the optically thick case,
which is preferred by photoionisation models,
lead to lateral expansion of the clouds. He termed the latter state
{\em pancake} clouds.
\citet{Mat86} then assessed the hydrodynamic stability of such pancake
clouds with the result that the lateral edges of a pancake cloud are
hydrodynamically unstable. Hence, the lateral flows persist and
destroy the cloud on a short timescale, comparable to one cloud orbit.
Also, the ram pressure by the
inter-cloud gas, which is probably pressure supported and moving in a
different way than the clouds, may compress the clouds in the
direction of relative motion and the increased pressure makes the
cloud\change{s} expand sideways \citep{Krolik88}, now with regard to the
direction of motion. These problems have lead to
the idea that clouds must reform or be re-injected
steadily. \citet{Krolik88} develops the idea of clouds forming by the
thermal instability from the hot inter-cloud gas. This idea has
however been rejected by \citet{MD90}: It would require very large
amplitude fluctuations of unusual type (high density, low
temperature). Additionally, a compression by a factor of 10,000 would
require an unusually small magnetic field, for the magnetic pressure not
to stop the collapse. \citet{MD90} further argue that the clouds would
be destroyed by the radiative shear mechanism before they could
contribute to the emission line profiles.
We have studied the radiative shearing mechanism
for dusty clouds in 2.5D hydrodynamic simulations \citep{SKB11}. In agreement with
\citet{Mat86} and \citet{MD90}, we find quick radiative shearing of
the cloud.

\change{As mentioned above, the stability considerations were for
  clouds accelerating due to the dominant radiation pressure. The
  problem is even more severe for bound
  clouds where gravity is comparable to the radiation force: The
  radiation force acts on the illuminated surface of the cloud, while
  every part of the cloud is uniformly subject to
  gravity which must dominate by definition for bound clouds. Such
  clouds are therefore compressed, and must consequently be stabilised
  by some internal pressure. Yet, radiation pressure and gravity act
  in one dimension, whereas the thermal pressure is isotropic. It is
  therefore not possible to stabilise illuminated bound clouds by
  thermal pressure alone. 
  }

In this context, we investigate the stability of magnetised and
gravitationally bound BLR clouds
-- initially close to an equilibrium orbit as calculated in paper~I. 
\change{ Magnetic fields have so far rarely been considered in BLR
  clouds. We therefore first review literature data for indications on
the relative magnitudes of gravitational, magnetic and radiative
forces.} 
(section~\ref{mclds}). 
\change{We then focus on
the effect of the magnetic field}, and simulate the evolution of
irradiated magneti\change{cally dominated bound} clouds.  We study the two-dimensional,
axisymmetric (2.5D), magnetohydrodynamic (MHD) evolution of isolated, 
initially optically thick clouds, including gravity,
rotation and radial radiation pressure via a simple equilibrium
photoionisation ansatz. We neglect self-gravity, viscosity and any
radiation source other than the central accretion disc. Because of the
small size of the clouds, our computational domain is small compared
to the full size of the BLR.  We describe technical details in
section~\ref{num}, the setup details in section~\ref{setup}, our results
in section~\ref{res} and discuss our findings in section~\ref{disc}. We
summarise our results in section~\ref{conc}.

\section{Magnetic fields, gravity and radiation pressure 
  in BLR clouds 
  -- indications from the literature}\label{mclds} 
% \change{
% Photoionisation models typically require a similar level of the radiation
% pressure on the clouds and the thermal pressure within. Yet,
% reverberation mapping results have placed the BLR closer to the AGN as 
% expected:}

\begin{figure}
\centering
\includegraphics[width=.47\textwidth]{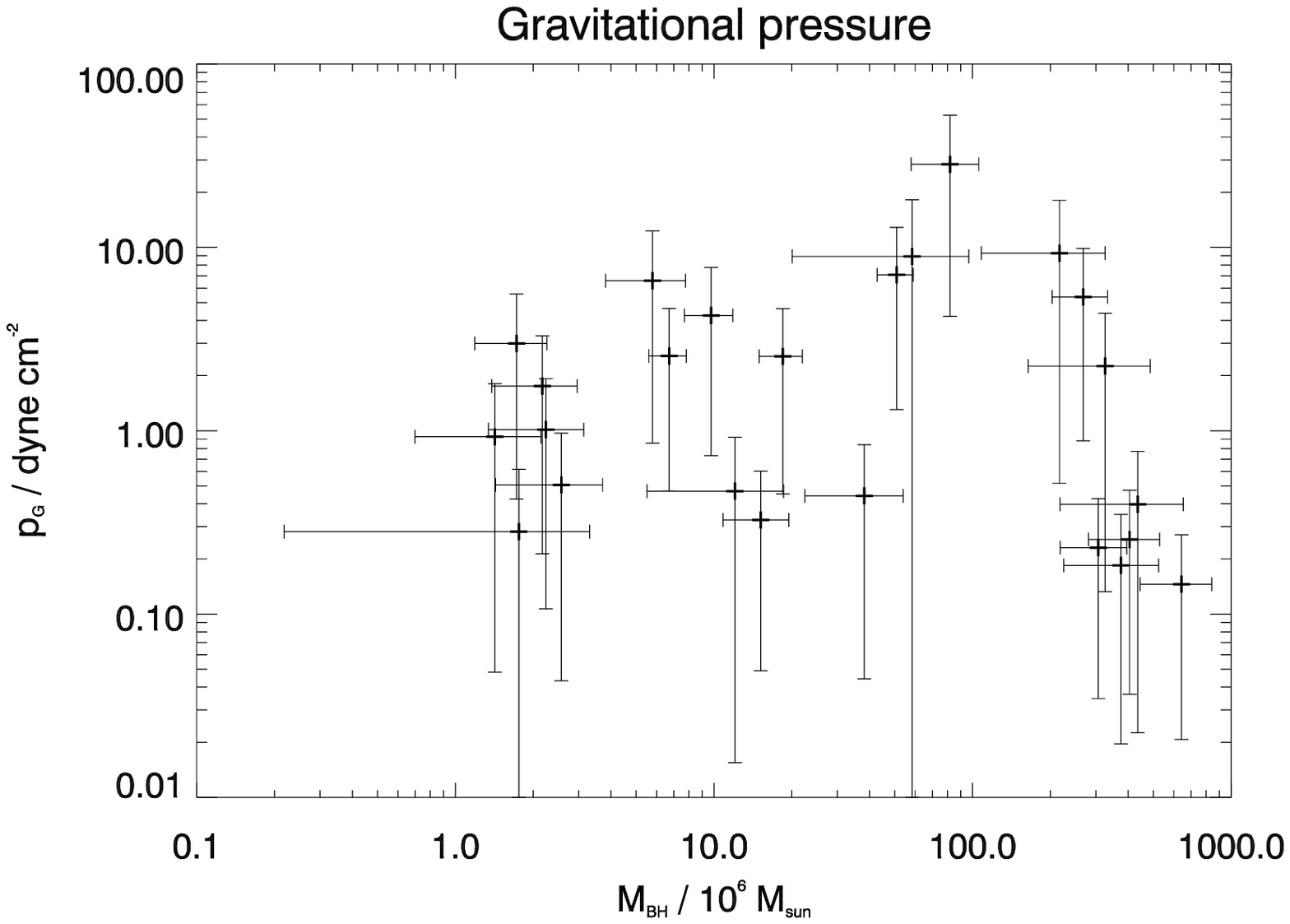}
\includegraphics[width=.47\textwidth]{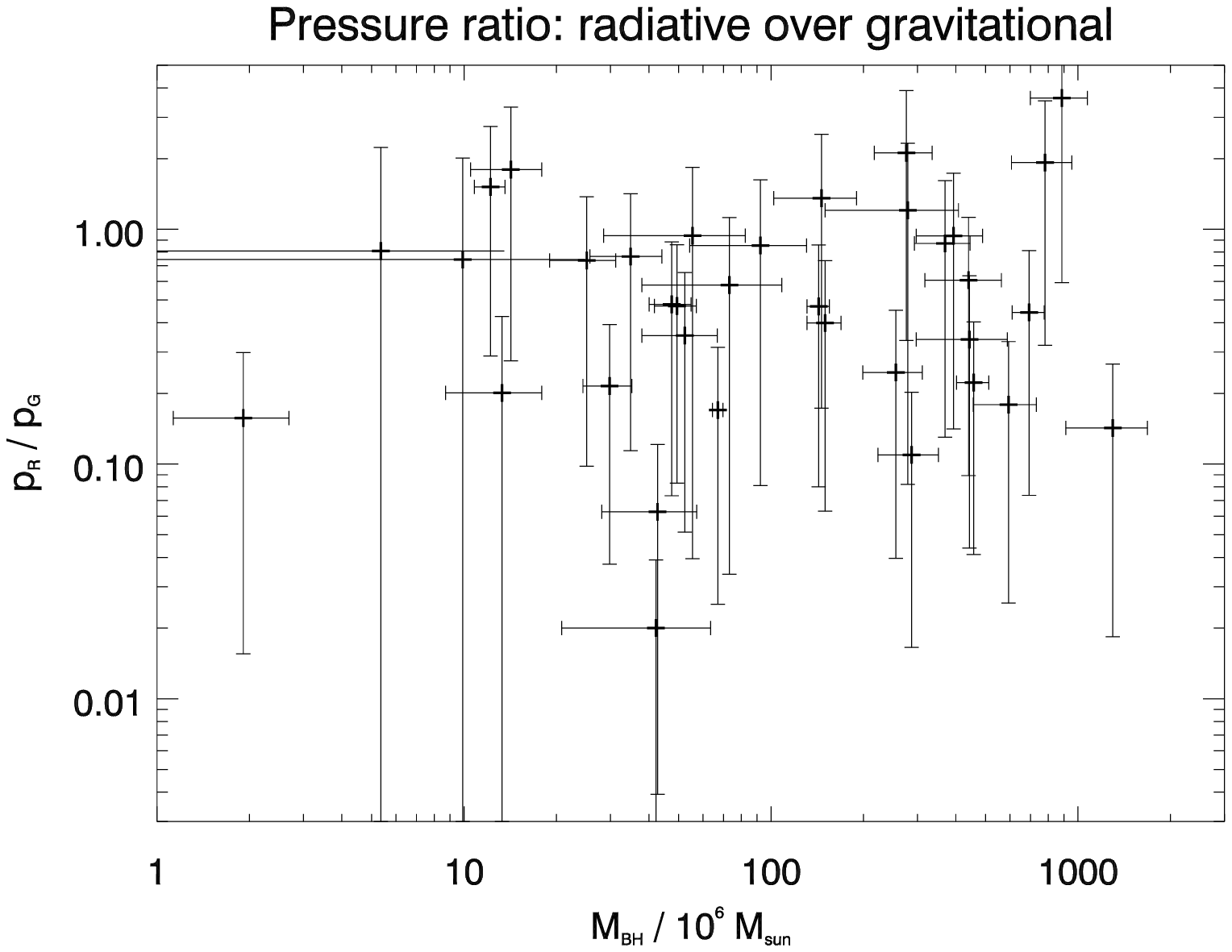}
\caption{\small  Radial pressure on BLR clouds. Top:
  Inwards pressure due to gravity over black hole mass (data of
  Table~\ref{rmdata}). Bottom: ratio
  of radiative to gravitational pressure \citep[all 35
  objects of][]{Peterea04a}. Different samples have been
  used for each plot, according to the availability of the data. 
  Error bars are propagated from the original publications (compare
  Table~\ref{rmdata}). Additionally, an error of 80~per cent on the
  assumed column density has been included. 
See text for details.}
\label{fig:rmdata}
\end{figure}

The typical pressure level imposed on the clouds by radiative, and
gravitational forces may be calculated from reverberation mapping (RM)
data
as follows: For this analysis we convert all forces to pressures,
dividing by the surface area of the cloud, assumed to be spherical in
this order of magnitude analysis.
The inwards pressure due to gravity on
the cloud is given by:
\eq{
p_\mathrm{G} = \frac{G M_\mathrm{BH} m_\mathrm{cld}}{r^2} \frac{1}{\pi
R_\mathrm{cld}^2} \, ,}
where $M_\mathrm{BH}$ is the black hole mass, $m_\mathrm{cld}$ the
mass of the cloud, $r$ the distance of the cloud from the black hole,
and $R_\mathrm{cld}$ the radius of the cloud.
In current reverberation mapping studies \citep[e.g.][]{Benea09}, the black
hole mass is determined as a function of the measured quantities of
the centroid time lag $\tau_\mathrm{cent}$ and the line dispersion of
the rest frame rms-spectra, $\sigma_\mathrm{line}$. Using these
variables, the gravitational pressure on the clouds may be expressed
as:
\eq{
p_\mathrm{G} = \frac{4}{3} f
\frac{R_\mathrm{cld}}{c\tau_\mathrm{cent}}
\rho_\mathrm{cld} \sigma_\mathrm{line}^2 \, .}
Here, $\rho_\mathrm{cld}$ is the density of the clouds, and $f=5.5$ is
the correction factor which ensures that RM-based black hole masses
agree with the $M_\mathrm{BH}$--$\sigma_*$ relation \citep{Onkea04}. We have
collected $\tau_\mathrm{cent}$ and $\sigma_\mathrm{line}$ from a
number of RM-studies where they were explicitly stated
(Table~\ref{rmdata}). Where available, we have included different
measurements of the same source at different epochs.
From this data we have calculated
$p_\mathrm{G}$ 
\rvtwo{using $R_\mathrm{cld}\rho_\mathrm{cld}=0.23$~g~cm$^{-2}$
  (corresponding to a hydrogen column density of $N_H=10^{23}$~cm$^{-2}$),} 
which is plotted against black hole mass in
Figure~\ref{fig:rmdata} (top). The values for $R_\mathrm{cld}$ and
$\rho_\mathrm{cld}$ are chosen to be in the range allowed from
photoionisation models (compare section~\ref{intro}). There seems to be no obvious
correlation. The mean of the gravitational pressure in the clouds
is~\rvtwo{3.63}~dyne/cm$^2$. Because the distribution seems to be more uniform in
log-space, we use in the following the median value, which is
\rvtwo{1.75~dyne/cm$^2$,} as a characteristic number.

The ratio between radiation pressure and gravitational pressure may
be expressed as:
\eq{ \label{prg_ratio}
\frac{p_\mathrm{R}}{p_\mathrm{G}} =
\frac{3}{16 \pi c G} \frac{1}{R_\mathrm{cld}\rho_\mathrm{cld}}
\frac{L_\mathrm{ion}}{M_\mathrm{BH}}\, ,}
where $L_\mathrm{ion}$ is that part of the AGN light that is absorbed by the cloud.
It is given by \citep[e.g.][]{Marcea08}
\rvtwo{$L_\mathrm{ion} = a b c L_{5100}$}, 
where $a=0.6$ is the ionising fraction of the bolometric
luminosity, $b=9$ is the bolometric correction and $L_{5100}$
represents $\lambda L_\lambda$ at 5100~\AA. 

\rvtwo{The factor $c$ accounts for geometrical and relativistic effects:
\citet{SM89} have shown that for frequencies up to about
  $\nu = 10^{15}$~Hz, the angular radiation pattern of 
  thin accretion discs around
  fast rotating black holes follow the classical $\cos(\theta)$-law
  (Lambert's law). For $\nu
  \gtrsim 10^{16}$~Hz, the emission is close to isotropic. For
Schwarzschild black holes, the transition occurs at somewhat higher
frequencies. Luminous AGN emit most of their energy in the innermost regions of
the accretion disc, in the blue -- ultraviolet part of the spectrum
(so-called ``big blue bump''), with a peak around $\nu_\mathrm{p}=
2\times 10^{15}$~Hz (e.g. \citeauthor*{HRH07} \citeyear{HRH07}).
If Lambert's law would apply and if the BLR would extend 20~degrees
above and below the equatorial plane, which would mean that about 10
per cent of the AGN's radiation would be intercepted, $c$ would
average to 0.3. For more rapidly spinning black holes or
accretion discs emitting at higher frequencies, $c$ might become as
large as unity. Also, if the BLR would not be located in the
equatorial plane of the accretion disc, $c$ would also tend towards
unity. Here, we adopt $c=1$ as possible, reasonable value.}

For these assumptions, we
have calculated the pressure ratio for the sample of \cite{Peterea04a}
(their complete Table~8). We have used this sample, because black hole
masses determined in the same way as for the sample above  as well as
continuum luminosities were available. We show a plot of the derived
pressure ratios against black hole mass in Figure~\ref{fig:rmdata}
(bottom). There is again no obvious correlation. The mean ratio of
radiation to gravitational pressure is \rvtwo{0.74}. Here, the distribution is
\change{also more uniform} in log-space, and therefore again we use the
median of the pressure ratios as a characteristic number, which
is \rvtwo{0.48}. 
These numbers depend of course crucially on the assumed column density.
  Photoionisation calculations constrain the column density to
  $N>2\times10^{22}$cm$^{-2}$ \citep{KK81}. We have used values for
  $R_\mathrm{cld}$ and $\rho_\mathrm{cld}$ corresponding to a
  \rvtwo{(hydrogen)} column
  of $N=10^{23}$cm$^{-2}$, consistent with the requirement from
  reverberation mapping studies that the clouds need to be optically
  thick (compare section~\ref{intro}, above). 
  \rvtwo{Allowing for the smallest column
  density consistent with the calculations of \citet{KK81} would shift
  the median of the pressure ratio to 2.39, with only 6 of 35 values still
  below unity. However,
\citet{Maiea10} and \citet{Risea11} have possibly observed BLR
  clouds in X-ray absorption and give values of a few times $10^{23}$ for
  the column of their clouds. With such a column density, the typical
  pressure ratio would clearly drop below unity. We have included a
  contribution of 80~per cent due to the uncertainties in the column
  density for the calculation of the error bars in
  Figure~\ref{fig:rmdata}. Another uncertainty
  is due to the accuracy of the black hole masses: As indication, the
  errors stated by \citet[their Table~13]{Benea09} are 43~per
  cent. Additionally, there may be a systematic error via $f$
  \citep{Onkea04}, and possibly corrections to radiation pressure
  \citep{Marcea08, Marcea09}. Because the pressure ratio values we
  find scatter around unity (Figure~\ref{fig:rmdata}), it is thus not
  possible to firmly conclude if the BLRs are typically
  gravitationally bound or otherwise.
 It is however remarkable that
  the pressure ratio distribution is bounded by a value close to unity for
  reasonable assumptions. In any case, gravitationally
  bound cloud models are clearly consistent with the data.}

%%%%%%%%%%%%%%%%%%%%%%%%%%%%%%%%%%%%%%%%%%%%%%%%%%%%
%%%%%%%%%%%%%%%%%%%%%%%%%%%%%%%%%%%%%%%%%%%%%%%%%%%%
\begin{table}
\begin{minipage}{0.8\textwidth}
\caption{Reverberation mapping data}\label{rmdata}
\begin{tabular}{lrrrr}\hline \hline
Object 
 & $\tau_\mathrm{cent}$\footnote{Centroid time lag.}
 & $\Delta \tau_\mathrm{cent} $\footnote{Error on centroid time lag.}
 & $\sigma_\mathrm{line} $\footnote{Line dispersion, rest frame
   rms-spectra.} 
 & $\Delta\sigma_\mathrm{line} $\footnote{Error on line dispersion.} \\
                         & (days)  &  (days)  & (km/s)  & (km/s) \\
\hline
\multicolumn{5}{c}{\citet{Benea09}} \\
Mrk142   		 & 2.74	&	0.75	&   859	& 102  \\
SBS1116+583A & 2.31	&	0.56	& 1528	& 184  \\
Arp151		 & 3.99	&	0.59	& 1252	&   46  \\
Mrk1310		 & 3.66	&	0.60	&   755	& 138  \\
Mrk202		 & 3.05	&	1.43	&   659	&   65  \\
NGC4253		 & 6.16	&	1.43	&   516	& 218  \\
NGC4748		 & 5.55	&	1.92	&   657	&   91  \\
NGC5548		 & 4.18	&	1.08	& 4270	& 292  \\
NGC6814		 & 6.64	&	0.89	& 1610	& 108  \\
\hline\multicolumn{5}{c}{\citet{Denea09a}} \\
NGC4051	       	 & 1.87	&	0.52 &   927	&   64  \\
\hline\multicolumn{5}{c}{\citet{Griea08}} \\
PG2130+099    & 22.9	&	4.7	& 1246	& 222  \\
\hline\multicolumn{5}{c}{\citet{Denea06}} \\
NGC4593	       	 & 3.73	&	0.75	& 1561	&   55  \\
\hline\multicolumn{5}{c}{\citet{Benea07}} \\
NGC5548		&   6.3	&	2.5	& 2939	& 768  \\
\hline\multicolumn{5}{c}{\citet{Benea06}} \\
NGC5548		&   6.6	&	1.0	& 2680	&   64  \\
\hline\multicolumn{5}{c}{\citet{Peterea04a}} \\
Mrk335	   	& 16.8	&	4.4	&   917	&   52  \\	
Mrk335		& 12.5	&	6.1	&   948	& 113  \\
PG0026+129	& 98.1	&    26.9	& 1961	& 135  \\
PG0026+129	&111.0	&    26.2	& 1773	& 285  \\
PG0052+251	&163.7	&    48.4	& 1913	&   85  \\
PG0052+251 	& 89.8	&    24.3	& 1783	&   86  \\
PG0052+251	& 81.6	&    17.7	& 2230	& 502  \\
Fairall9		& 17.4	&	3.8	& 3787	& 197  \\
Fairall9		& 29.6	&    13.7	& 3201	& 285  \\
Fairall9		& 11.9	&      5.7	& 4120	& 308  \\
\hline\hline
\end{tabular}
\end{minipage}
\end{table}

The typical radiation pressure on BLR clouds is given by the
characteristic gravitational pressure times the typical pressure
ratio, and using the \rvtwo{fiducial} numbers derived above:
\eql{prad}{\bar{p}_\mathrm{R} = 0.84\,\mathrm{dyne\,cm}^{-2}\, .}
This number exceeds the \rvtwo{upper bound for the} 
thermal pressure derived from 
\change{standard} photoionisation
models (compare above) by \rvtwo{ a factor of a few}. It is
therefore non-negligible. Even, if one takes the lower bounds of the
distributions of $p_\mathrm{G}$ and $p_\mathrm{R}/p_\mathrm{G}$, the
resulting radiation pressure \rvtwo{is well within the range of inferred
values for the}
thermal pressure. 

The fact that the
 reverberation mapping results imply small BLR radii and therefore high
radiative fluxes has already been noted by several authors
\citep{Pet88,Ferlea92,LC07}. It would seem to imply that the photoionisation
parameter is high. Adopting $\Xi=2.3 p_\mathrm{R}/p_\mathrm{T}$
as photoionisation parameter \citep{KMKT81}, we arrive at a
characteristic value of $\Xi=14$, taking the upper bound of the thermal
pressure from the photoionisation models, and ten times higher for the
central value. \citet{KMKT81} point out two
arguments to constrain $\Xi$: First, pressure balance with a
surrounding hot phase implies $1/3 \lesssim \Xi \lesssim 10$.
Second, fitting the line ratios, requires 
$0.3 \lesssim \Xi \lesssim 2$ \citep{KK81}. 
\rvtwo{Additionally, very high ionisation parameters lead of course to highly
ionised gas, which would be unable to produce the observed emission
lines. Yet, the uncertainties in the column density, geometrical and
relativistic correction factors and black hole masses are
considerable. Thus, while the ratio of radiation pressure to thermal
pressure comes
out a little high, which might point e.g. to a lower
value of our factor $c$ (compare above) or cloud densities which are
generally rather around $10^{11}$~cm$^{-3}$, the data still seems to
be consistent.}

In summary, the data suggests that gravitational and radiation
pressure are both important and at least comparable to
the thermal pressure. Gravitationally bound clouds are are consistent with all
the available data.

We have argued above that thermal pressure, due to
its isotropic nature may not stabilise the cloud against the
uni-directional opposing forces due to gravity and radiation. 
Magnetic forces are not isotropic. Their direction depends on the
geometry of the magnetic field. It is therefore natural to ask if a
magnetic field configuration may be found that is able to support an
illuminated bound cloud. If the magnetic field should be at all
capable to contribute, its energy density should be at least comparable to the
radiation pressure. Both, magnetic pressure and magnetic tension are
of the same order of magnitude as the magnetic energy density for
many, not highly symmetric configurations.
\rvtwo{Accordingly, the magnetic field strength required for the
  magnetic energy density to match our reference radiation pressure
  (equation~(\ref{prad})) is 5~G. Taking into account the accuracy of
  the measurements, as well as the width of the distribution of
  pressures (Figure~\ref{rmdata}), the true indicative value for any
  given BLR should be within a factor of ten of this number.}

\change{
Guided by these considerations, we make the following assumptions
for our magnetohydrodynamic stability analysis:
\begin{enumerate}
\item The clouds are gravitationally bound.
\item The magnitude of the radiation pressure is significant compared
  to gravity.
\item The magnetic pressure is at least comparable to the radiation pressure.
\item The thermal pressure of the cloud is much smaller than both the radiation
  pressure and the magnetic pressure.
\end{enumerate}
The last assumption is mainly for methodological reasons, in order to
isolate the effects of the magnetic field.}
In the following we present magnetohydrodynamic simulations of clouds
set up 
\change{according to the standard cloud properties outlined in
  section~\ref{intro}, and the above assumptions}, 
and investigate their stability numerically.

\section{Numerical model}\label{num}
The basic code we use is the 3D~MHD code {\sc
  NIRVANA} \citep{ZY97}. For short, it conserves mass,
momentum and internal energy in the advection step, interpolates
the fluxes with van
Leer's formula \citep[second order accurate]{vL77}, and uses the
constrained transport method to keep the magnetic field divergence free.

In order to include the effects of photoionisation and electron scattering,
we augmented the MHD equations by an equation for the radial
radiative transfer for the photon flux $S$:
\begin{equation}
\frac{\partial S}{\partial r}= 
%-\alpha_\mathrm{rec} (n\chi)^2
%\frac{d\,n_\mathrm{H\,II}{d\,t}
-\dot{n}_\mathrm{H\,I}
-\left(\frac{2}{r} +\sigma_\mathrm{T} n \chi\right) S\, ,
\end{equation}
where the first term on the right-hand side is due to
photoionisation, and the second one describes the geometrical effect
and electron scattering, with the Thomson cross section $\sigma_\mathrm{T}$.
In equilibrium, the ionisation rate $\dot{n}_\mathrm{H\,I}$ is given
by the number of recombinations:
\begin{equation}
\dot{n}_\mathrm{H\,I}= \alpha_\mathrm{rec} (n \chi)^2 \, ,
\end{equation}
where $n$ is the total number density of ionised and neutral hydrogen
atoms. 
We assume a recombination coefficient of
$\alpha_\mathrm{rec}=5 \times 10^{-13}$~cm$^3$~s$^{-1}$
\citep[e.g.][]{ADU}, kept constant for simplicity. 
The ionisation fraction $\chi$ is given by:
\begin{eqnarray}
y&=& \frac{\sigma_\mathrm{phot} S}{\alpha_\mathrm{rec} n} \\
\chi&=&-\frac{y}{2} + \sqrt{\left(\frac{y}{2}\right)^2+y} \, .
\end{eqnarray} 
We assume a photoionisation cross section of
$\sigma_\mathrm{phot}=6.3\times 10^{-18}/(\alpha+3.5)$~cm$^2$, averaged over a
spectrum $\nu^{-\alpha}$ with $\alpha=2$. This leads to a local
radiative acceleration of:
\begin{equation}
a_\mathrm{rad}=(\dot{n}_\mathrm{H\,I}+\chi n \sigma_\mathrm{T} S)
h\bar{\nu}/\rho c \, ,
\end{equation}
where $\rho$ denotes the mass density and $c$ the speed of light. We
assume an average photon energy \rvtwo{$ h\bar{\nu}= 1.007 h\,\nu_0$, with the
hydrogen ionisation potential $h\,\nu_0=13.6$~eV}, for a steeply
falling spectrum. The radiative heating is not taken into account
explicitly.
\rvtwo{We set however a lower limit to the radiative cooling at
  1000~K, which implies a heat term that balances cooling in such a
  way as to reach the minimum temperature. We}
\change{do not attempt to model the temperatures within the
  clouds correctly. This is justified, because we study the evolution
  of magnetically dominated clouds, and even if photoionisation
  heating was taken into account, the thermal pressure would remain
  negligible in our setup.
We set the minimum temperature below the one expected from
photoionisation equilibrium, in order to investigate if \rvtwo{shock
  heating} can also heat the clouds to a temperature comparable to the
photoionisation temperature.} 

Equilibrium photoionisation is a good assumption,
because the ionisation front would pass the clouds quickly ($\lesssim$
hour), and the thermal pressure imbalance induced by the passing of
the ionisation front would be insignificant compared to the magnetic
pressure. The clouds we consider below are about halfway
ionised. Thomson scattering dominates in the inter-cloud gas, hydrogen
photoionisation within the clouds, which also have a still significant
contribution by Thomson scattering. 
This model was chosen with the aim to be as simple
as possible, while capturing the essential physics (we are interested
in the radiative
acceleration). We have varied
the parameters within reasonable limits without a qualitative change
of the results.
We adopt an equilibrium solar metallicity cooling curve \citep{SD93},
extrapolated to lower temperatures as described in \citet{KA07}.
Finally, we take into account the gravitational acceleration $g$ 
due to a central point source (the
SMBH).

These effects are implemented as source terms into the system of
MHD equations:
\begin{eqnarray}
\frac{\partial \rho}{\partial t} + \nabla \cdot \left( \rho {\bf v}\right)&
 = & 0 \\
\frac{\partial \rho {\bf v}}{\partial t} + \nabla \cdot\left( \rho {\bf v} 
{\bf v} \right) & = &
- \nabla p +\frac{1}{4\pi} ({\bf B} \cdot\nabla){\bf B} -
\frac{1}{8\pi}\nabla{\bf B}^2 \nonumber\\
& &+ \rho (a_\mathrm{rad}-g)\\
\frac{\partial e}{\partial t} + \nabla \cdot \left(e {\bf v} \right) & = &
- p \; \nabla \cdot {\bf v} - {\cal C}\\
\frac{\partial {\bf B}}{\partial t} &=& \nabla \times ({\bf v}\times {\bf B}) 
\label{ie} \, , 
\end{eqnarray}
where $\rho$ denotes the density, $e$ internal energy density, 
${\bf v}$ velocity,
${\cal C}$ the radiative cooling and $p=(\gamma -1) e$ the pressure.

The problem we are trying to address here has some particular
requirements: We have verified that angular
momentum is conserved to high accuracy during our simulation
runs. \citet{KA07} have shown that energy is conserved well in
multi-phase setups up to density contrasts of about $10^7$. Also in
the simulations presented here, we have checked that the total energy
is reasonably well conserved. We are however limited in density
contrast to about $10^4$.

\section{Simulation setup}\label{setup}
\begin{table*}
\begin{minipage}{0.8\textwidth}
\caption{Simulation parameters}\label{simpars}
\begin{tabular}{llrrrrrrrrr}\hline \hline
  Label & comment 
& $r_\mathrm{in}$\footnote{Inner boundary of the computational domain in $r$ direction
    in $\mu$pc. Note that the origin is not within the computational domain.}
& $r_\mathrm{out}$\footnote{Outer boundary of the computational domain
  in $r$ direction in $\mu$pc. }
& $\Delta \theta$\footnote{Size of computational domain in
 $\theta$ direction, symmetrical above and below the equator in units of $10^{-4}\pi$.} 
& $\Delta z$\footnote{Corresponding approximate domain size
 in vertical direction in $\mu$pc.} 
& $R_\mathrm{cld}$\footnote{Radius of the cloud in $\mu$pc.} 
& $n_r$\footnote{Number of cells in radial direction.} 
& $n_\theta$\footnote{Number of cells in meridional direction.} 
&$\delta r$\footnote{Uniform resolution in $\mu$pc/cell.} 
& $V$\footnote{Initial cloud orbital velocity as a fraction of the
  Kepler velocity in per cent.}\\  % These numbers are wrong in NV.out
                                % files due to a forgotten root
                                % function in setmod.c. I have
                                % corrected them here by running the
                                % code startup again with the
                                % corrected sqrt. mghk@mpe, 4March2011
\hline
R34 & eq-cld          &9100& 9250 & 32 & 92    & 3 & 750 & 460 & 0.2 &85\\
R35 & neq-cld        &9120& 9320 & 32 & 93  & 3 & 1000 & 460 & 0.2 &85 \\
R36 & R34-hires     &9100& 9250 & 32 & 92  & 3 & 1500 & 920 & 0.1 &85\\
R47 & R34-big-cld &9000& 9300 & 64 & 188 & 6 & 750 & 460 & 0.4 & 94\\
%R49 & helical       &9150 & 9250 & 32 & 92  & 3 & 1000 & 920 & 0.1 & 85\\
R60 & helical        &9150& 9250 & 32 & 92 & 3 & 1000 &  920 & 0.1 & 85\\
R61 & helical-hires      &9150& 9250 & 32 & 92 & 3 & 1500 & 1380 & 0.067 & 85 \\ 
R62 & helical-lores      &9150& 9250 & 32 & 92 & 3 &   500 &   460 & 0.2 & 85 \\
R63 & helical-big         &9100& 9300 & 64 & 188 & 3 & 2000 & 1840 & 0.1 & 85 \\
\hline \hline
\end{tabular}
\end{minipage}
\end{table*}

We simulate isolated clouds in axisymmetry. 
\change{The initial geometry is toroidal with a circular shape in the
  meridional section.}
The
basic simulation parameters are typical for BLRs and are summarised in
table~\ref{simpars}. The cloud is always positioned at a distance of
9200~{$\mu$}pc\footnote{\change{$10^{-6}$ parsec}} from the central SMBH. We resolve
\change{the cloud} diameter \change{of 6~{$\mu$}pc} with
30~cells in standard runs, \change{and up to 90~cells in the highest}
resolution ones. We set the initial cloud density to 
\change{$4 \times10^{10} m_\mathrm{p}$~cm$^{-3}$}, 
where $m_\mathrm{p}$ is the proton mass.
This results in a central \rvtwo{hydrogen column density of
$5\times10^{23}$cm$^{-2}$}, except for run R47, which has twice this
value.

%%%%%%%%%%%%%%%%%%%%%%%%%%%%%%%%%%%%%%%%%%%%%%%%%%%%
\begin{figure}
\centering
\includegraphics[width=.47\textwidth]{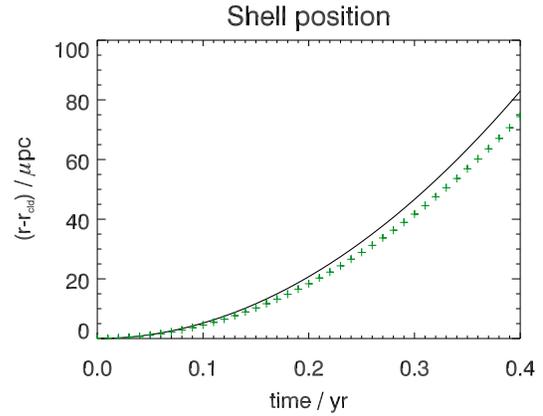}
\caption{\small  Radial position of the centre of mass of a 1D test
  cloud (green plus-symbols)
over time compared to the analytic prediction (solid black line) 
for an optically thick cloud.}
\label{fig:test}
\end{figure}
%%%%%%%%%%%%%%%%%%%%%%%%%%%%%%%%%%%%%%%%%%%%%%%%%%%%
 The setup follows essentially the scenario of
paper~I: We assume a hot atmosphere in approximate hydrostatic
equilibrium, where the pressure follows a power law $p(x)=p_0
x^{-s}$ with $x=r/r_\mathrm{cld}$ and $s=2$. Here, $r_\mathrm{cld}$ denotes the
radial position of the cloud (distance from the SMBH) 
and $p_0= GM_\mathrm{BH}\rho_0(1-\Gamma)/r_\mathrm{cld}s$ is what the
inter-cloud thermal pressure would be at that location, with
$\rho_0=10^8 m_\mathrm{p}$~cm$^{-3}$ and the Eddington factor $\Gamma=0.1$
($l$ in paper~I). Here, we take radiation pressure due to
Thomson scattering into account. $G$ is the gravitational constant
and $M_\mathrm{BH}$ the black hole mass, which is generally assumed
to be $10^8$ solar masses. The density is given by
$\rho(r)=\rho_0 x^{1-s}$. The magnetic field (compare below) in the
inter-cloud medium is perpendicular to the computational domain and
contributes initially about ten per cent to the total pressure, above
hydrostatic equilibrium. Yet, because of the cooling the atmosphere
still develops an inwards flow over the simulation time.

The pressure level might appear high, yielding an inter-cloud
temperature of order $10^9$~K. This is however necessary since we
would like to test here the effect of dynamically relevant magnetic
fields primarily inside the clouds. In order to yield a smaller
thermal inter-cloud pressure, and therefore a temperature of order the
Compton temperature, which one might expect, one would have to assume
a much stronger magnetic field also in the inter-cloud medium. We
defer the investigation of this situation to future work.

According to paper~I, force equilibrium is reached for the cloud, if
the rotational velocity in Kepler units is given by:
\begin{equation}\label{eq:veq}
V_\mathrm{eq}^2 = 1- \frac{3\Gamma}{2\sigma_{T}N} \, ,
\end{equation}
where \rvtwo{$N=2R_\mathrm{cld}\rho_\mathrm{cld}/m_\mathrm{p}$ is a measure of the column density of the cloud.}
In paper~I, we have found that for high column
densities, the clouds should be in a minimum of the effective potential,
and therefore remain stable at their radial position. For low column
densities, the clouds should be either ejected, or be sent inwards on
an excentric orbit. For our pressure profile, the critical rotation
velocity is 75~per~cent of the Kepler 
velocity\footnote{Note that eq. (6) of paper~I contains a mistake. The
correct formula for the critical rotation velocity should read:
$V_\mathrm{c}^2=(1+3/2s)^{-1}$, with the pressure power law index
$s=2$ in our case.}.
All our clouds are set up on the stable
branch of the equilibrium curve (compare Table~\ref{simpars}). 
We found experimentally that we require a rotational velocity of
102~per cent of the value given in Equation~\ref{eq:veq}
to ensure initial force equilibrium. 
The difference is due to the small amount 
of Thomson scattering in
the ambient medium on the way to the cloud, which reduces the amount
of radiation that actually arrives at the cloud. 
% The pathlength varies with
% the size of the computational domain and also the vertical position
% since the cloud is initially circular in the meridional plane. Also,
% the amount of radiation absorbed by the cloud varies vertically due to
% the varying column density, highest through the centre of the cloud
% and vanishing towards its edges. 

For runs R34, R36 and R47, we use an azimuthal magnetic field, only. The plasma
$\beta=8\pi p/B^2$ is initially set to 10 outside the cloud. Inside
the cloud, we set the temperature to the minimum one (1000~K), and
adjust the azimuthal magnetic field to yield pressure balance. This
results in a magnetic field about three times stronger than in the
inter-cloud medium, and a $\beta=10^{-4}$.
Run R35 uses a uniform field strength inside and outside the cloud
again with $\beta=10$ in the inter-cloud medium. This cloud is
therefore initially underpressured.
Runs R60, R61, R62 and R63 have an additional poloidal field 
(in the cloud, only). We have again
$\beta=10$ in the inter-cloud medium. Within the cloud, the azimuthal
field declines with distance from the cloud centre as a Gaussian:
\eq{B_\phi=B_{\phi,\mathrm{ic}}
  (1+3\, \mathrm{exp}(-10R^2/R_\mathrm{cld}^2))\, ,}
where $B_{\phi,\mathrm{ic}}$ is the azimuthal inter-cloud field, and
capital $R$ refers to the distance from the cloud
centre. $R_\mathrm{cld}$ is the cloud radius.
The poloidal field is set up as closed field loop in a given
meridional plane.  The 3D structure would be helical. We initialise the
radial component by
\eq{B_r=\frac{\alpha r_\mathrm{cld}}{r \sin \theta} \frac{\partial}{\partial
    \theta}B_\phi \sin\theta\, .}
We have tried different values for $\alpha$. The one we report here is
$\alpha=2\times10^{-5}$. 
\rvtwo{For this case, the initial peak values of the toroidal and poloidal
  field components are 103~G and 15~G, respectively.}
The magnetic field strength is for all runs \rvtwo{generally} around
30~G (the larger initial field in the runs with helical magnetic field
decays quickly to this value), to ensure magnetically dominated clouds, following the
considerations in section~\ref{mclds}.

All boundary conditions are set to zero gradient for all variables.
We believe that the best test for our radiation pressure module is the
approximate
stability of the hydrostatic halo, mentioned above. We have also
checked 1D-cloud acceleration without gravity (Figure~\ref{fig:test}).
In this test, the cloud accelerates a bit more slowly than expected
from simple radiative acceleration of an optically thick cloud. This
is, because the cloud also has to work against the ram
pressure of the ambient medium.

\section{Results}\label{res}
%%%%%%%%%%%%%%%%%%%%%%%%%%%%%%%%%%%%%%%%%%%%%%%%%%%%
\begin{figure*}
\centering
\includegraphics[width=\textwidth]{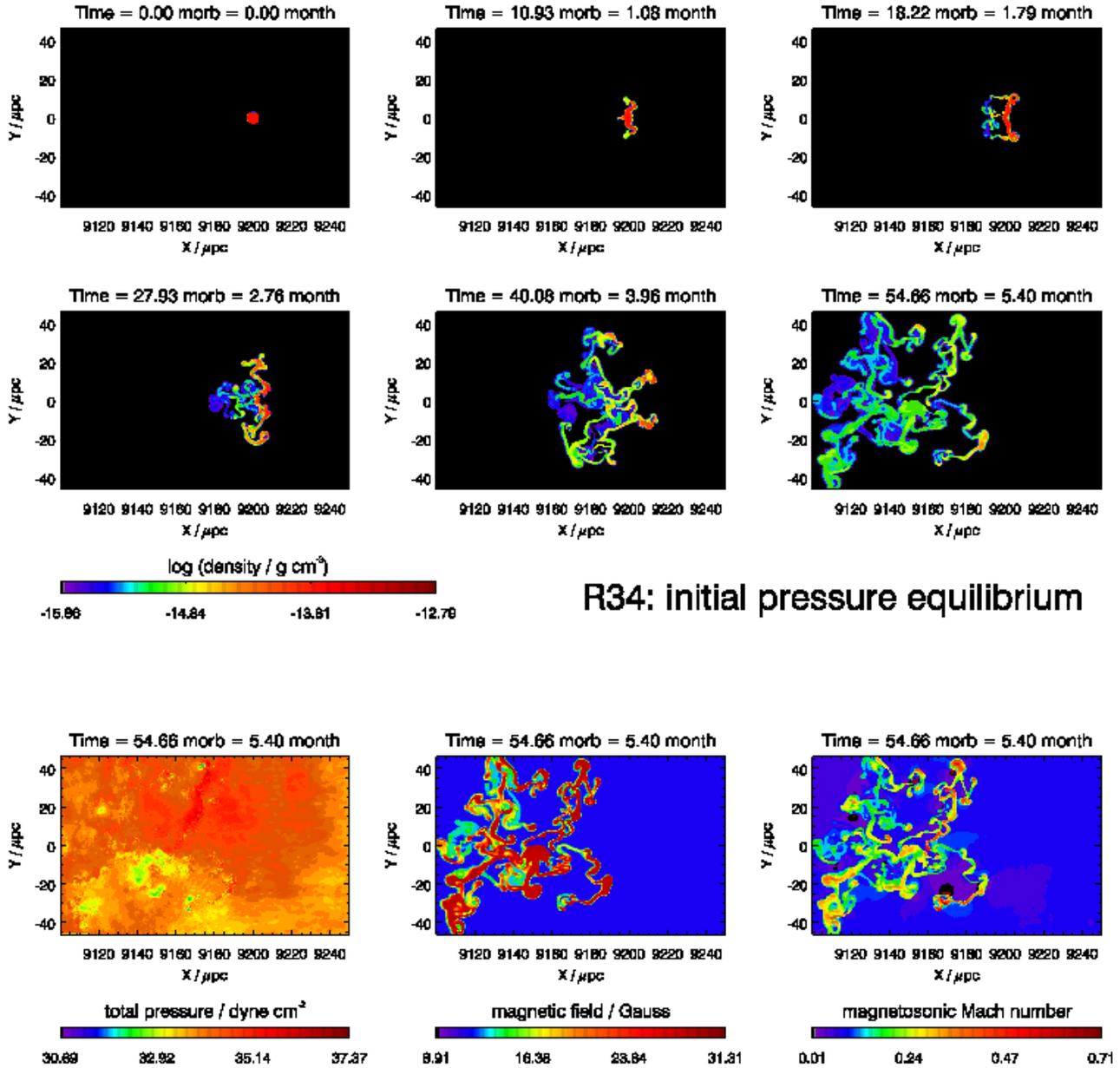}
\caption{\small  Evolution of run R34. The top six images show the
  logarithm of the density for different snapshot times, indicated in
  milli orbits (morb) and months on
  the individual images. The bottom three images show for the final
  snapshot from left to
  right: Total pressure, magnetic field strength and magnetosonic Mach
number for the velocity component in the meridional plane. In the cold
cloud gas, the magnetosonic speed is almost the same as the
Alfv\'en speed.}
\label{fig:R34o}
\end{figure*}
%%%%%%%%%%%%%%%%%%%%%%%%%%%%%%%%%%%%%%%%%%%%%%%%%%%%
\begin{figure*}
\centering
\includegraphics[width=\textwidth]{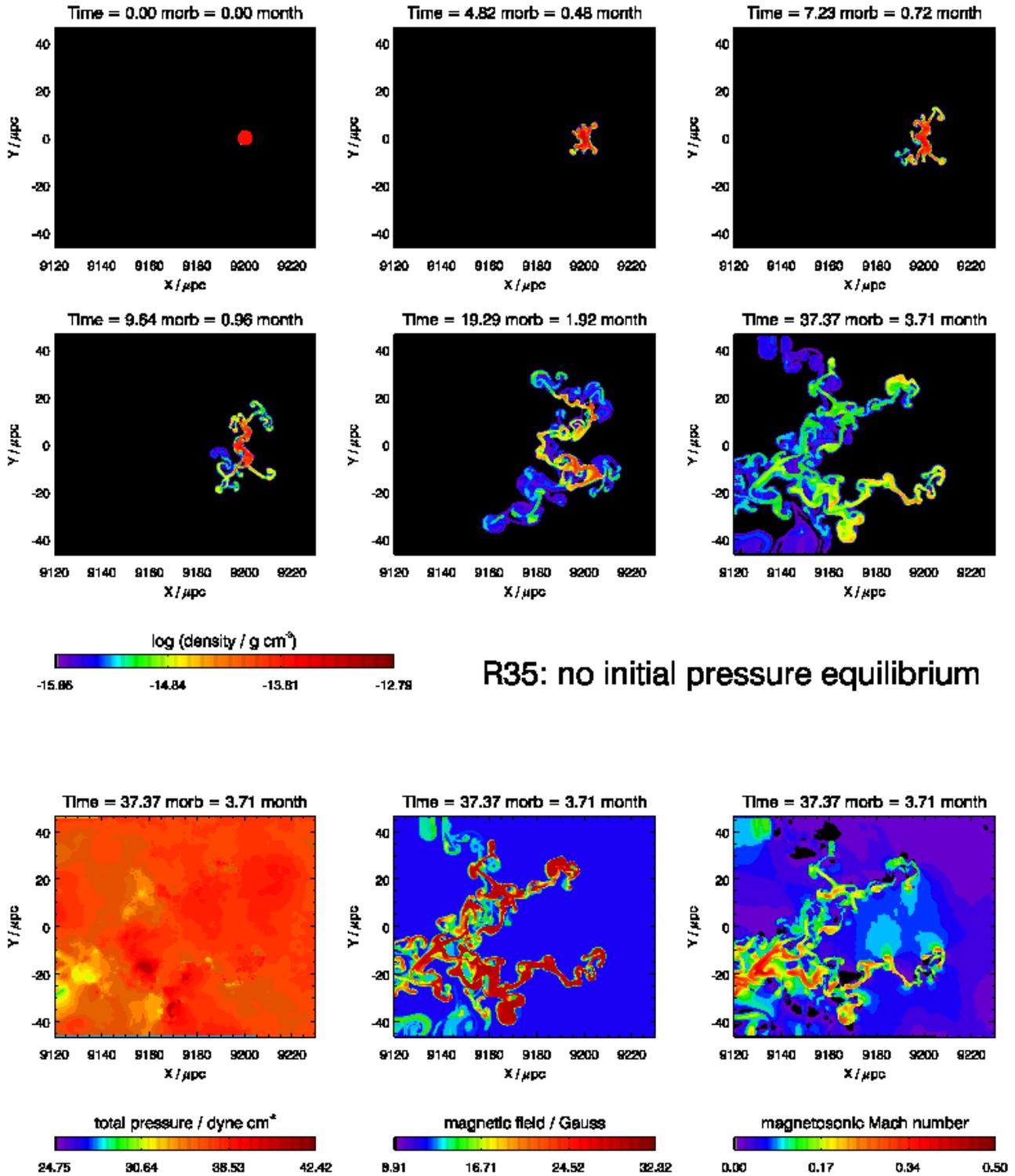}
\caption{\small  Same as Figure~\ref{fig:R34o} for run R35.}
\label{fig:R35o}
\end{figure*}
%%%%%%%%%%%%%%%%%%%%%%%%%%%%%%%%%%%%%%%%%%%%%%%%%%%%
\begin{figure*}
\centering
\includegraphics[width=\textwidth]{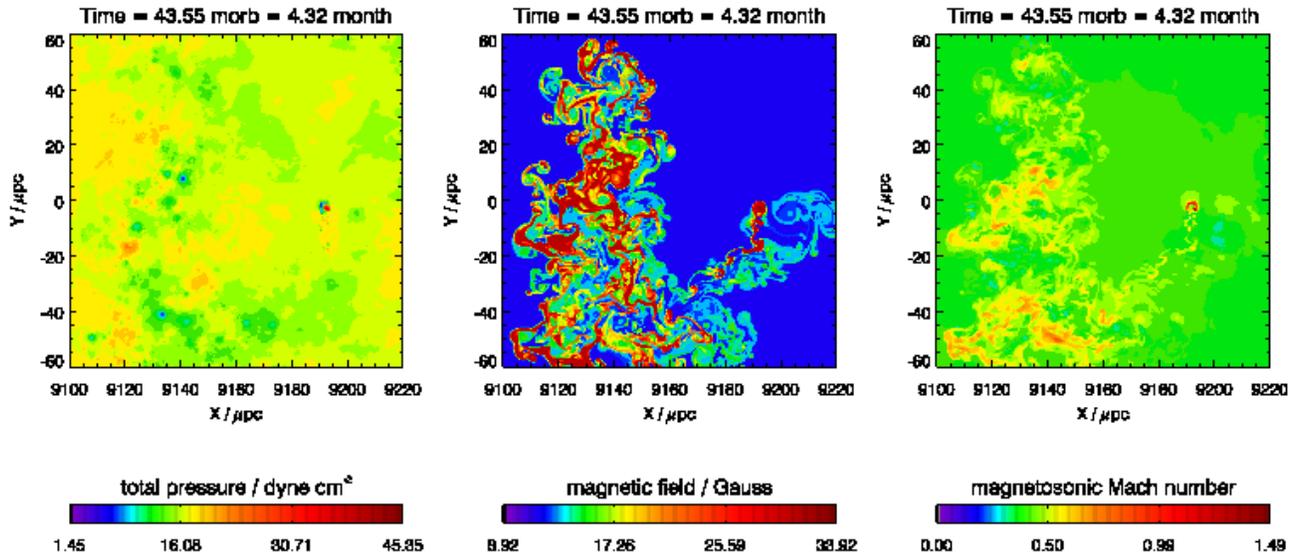}
\caption{\small  Same as Figure~\ref{fig:R34o} for run R63. \rvtwo{Only part
  of the computational domain is shown.}}
\label{fig:R63o}
\end{figure*}
%%%%%%%%%%%%%%%%%%%%%%%%%%%%%%%%%%%%%%%%%%%%%%%%%%%%
\begin{figure*}
\centering
\includegraphics[width=.49\textwidth]{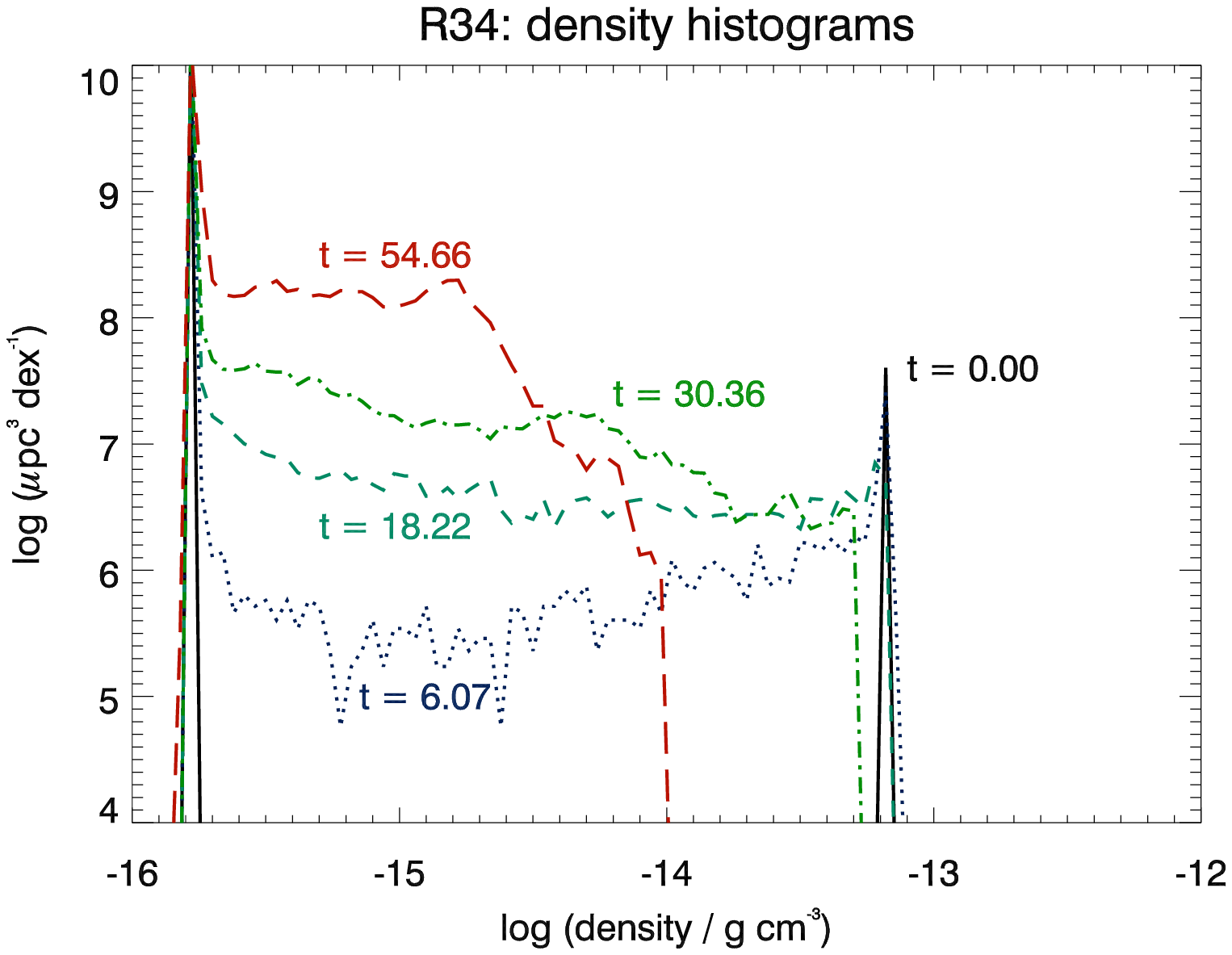}
\includegraphics[width=.49\textwidth]{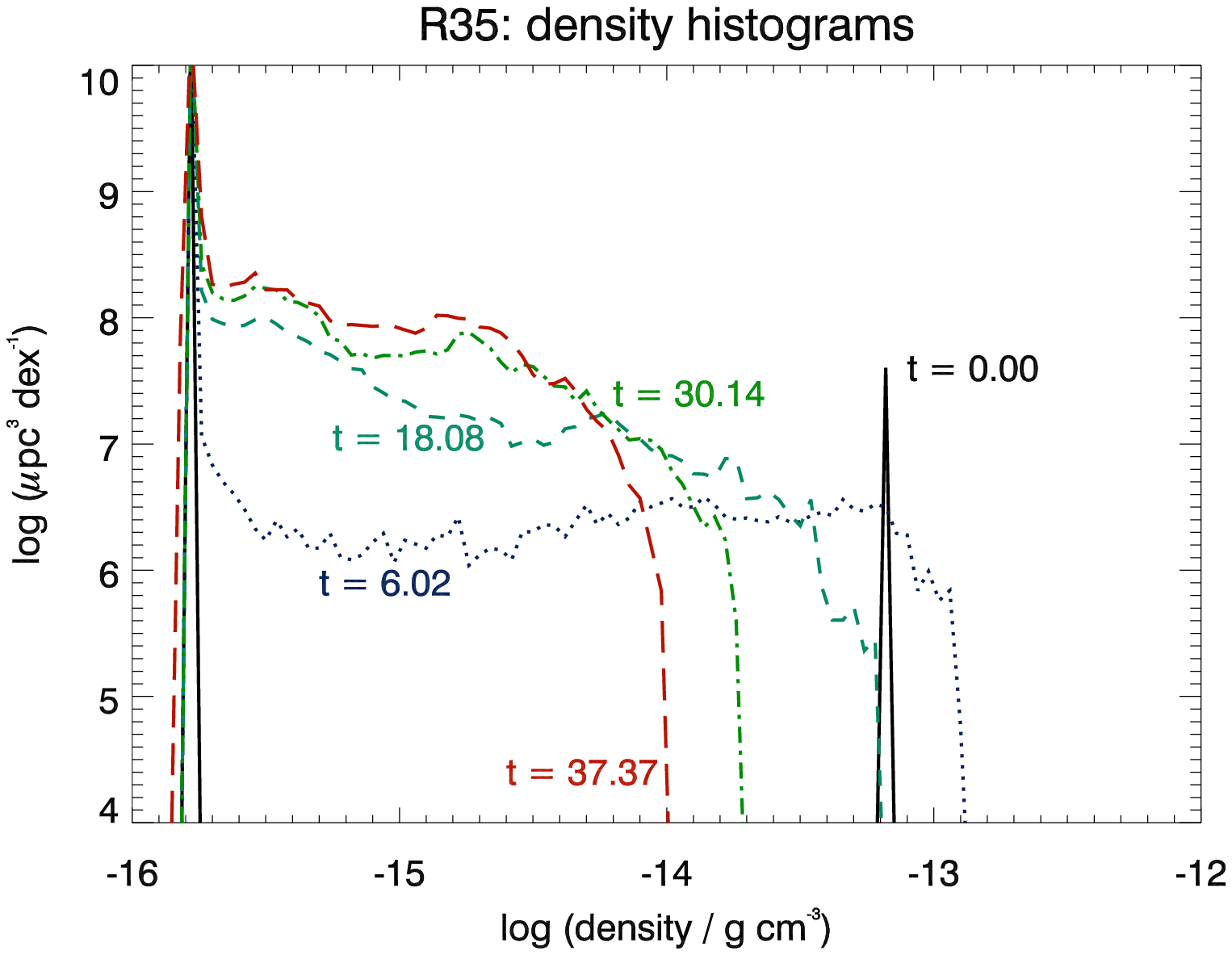}
\includegraphics[width=.49\textwidth]{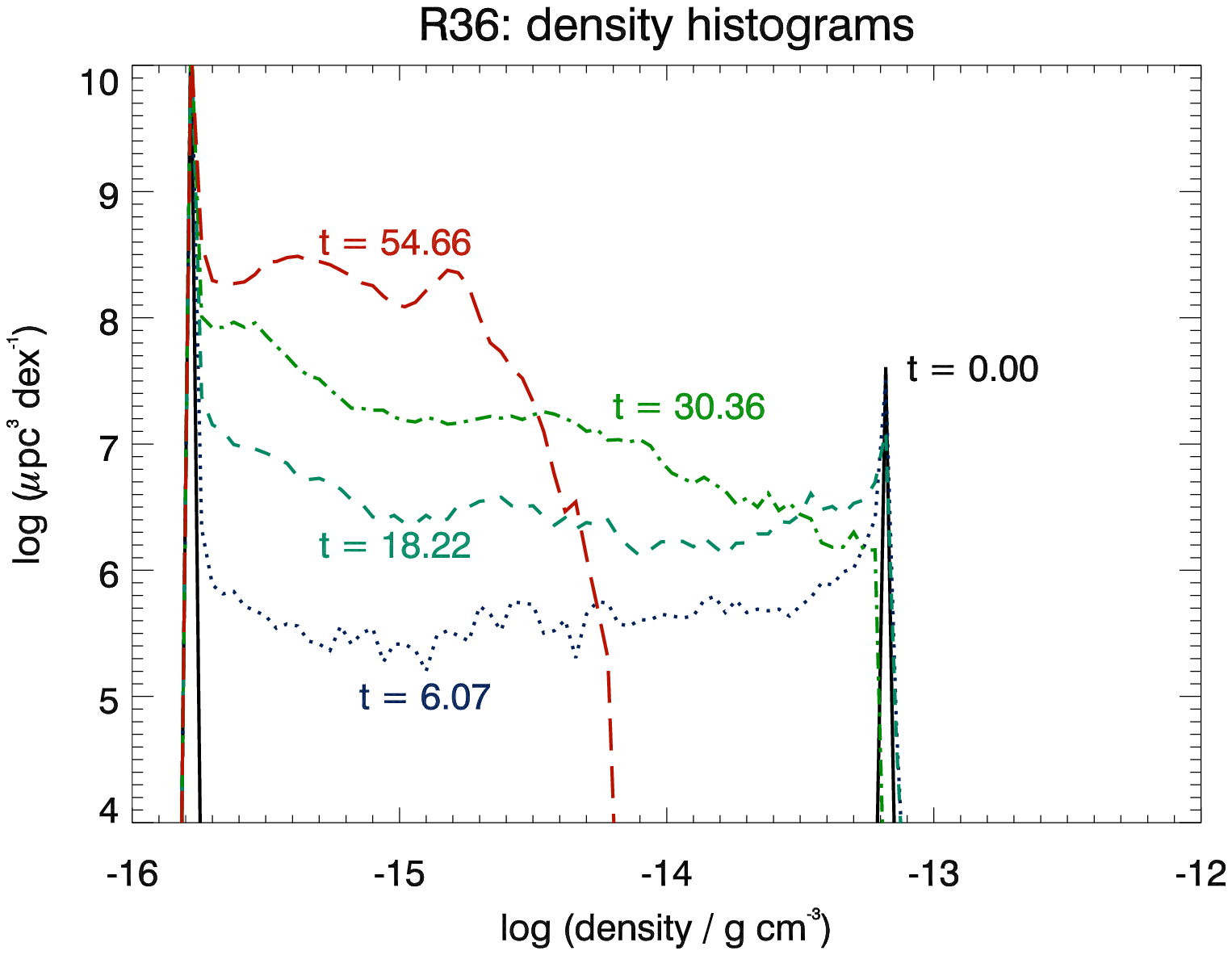}
\includegraphics[width=.49\textwidth]{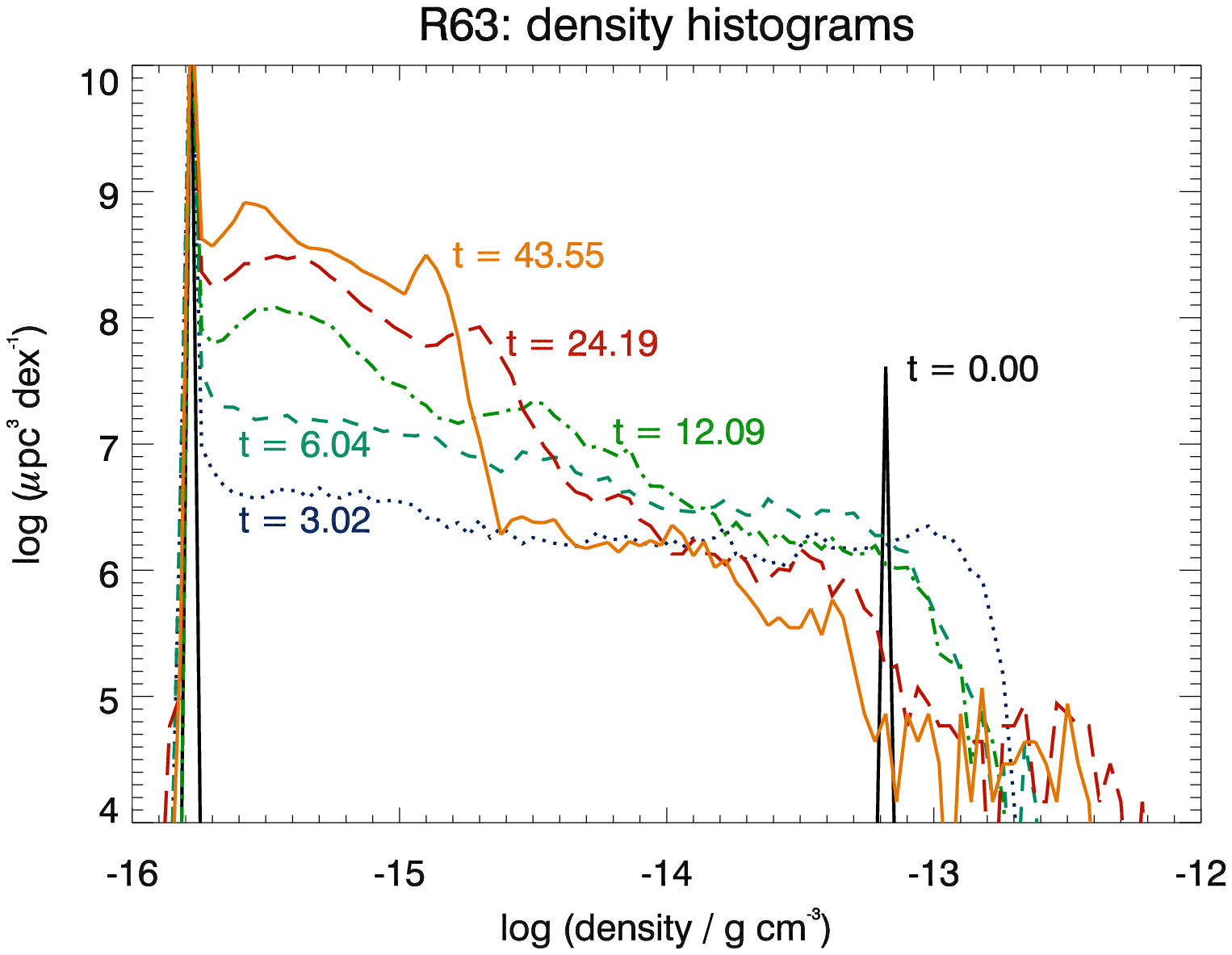}
\caption{\small  Density histograms for various runs, indicated in the
  individual titles and snapshot times (order: rder: solid black,
  dotted blue, dashed green, dot-dashed bright green, dashed red, solid orange),
  indicated in milli-orbits (morb) next to the
  respective lines. Mixing generally moves the upper density cutoff
  towards lower density (leftwards) for later times. The exception is
  run~R63 for which the histogram converges at $t\approx 9$~morb
  and also continuously 
  extends towards densities higher than the initial cloud density.}
\label{fig:dhistos}
\end{figure*}
%%%%%%%%%%%%%%%%%%%%%%%%%%%%%%%%%%%%%%%%%%%%%%%%%%%%

\subsection{General dynamics}\label{res-gd}

A general overview of the evolution of our simulated magnetised clouds
is shown in Figures~\ref{fig:R34o},~\ref{fig:R35o} and
~\ref{fig:R63o}. Movies are provided with the online version.

\subsubsection{Pressure equilibrium}
We first describe the evolution for run~R34 (Figure~\ref{fig:R34o}), 
set up in initial total
pressure equilibrium with an azimuthal magnetic field, only. The cloud
is stationary at its radial position due to the matched radiative and
centrifugal forces on the one hand, and the gravitational force on the
other hand. This leads to a radial compression of the cloud
(10.93 morb ($=$~milli-orbits)). At the same time, the magnetic pressure in the cloud
increases. As there is no opposing force vertically, the cloud expands
upwards and downwards, much like an open tube of toothpaste squeezed
in the middle. 
This evolution matches the picture that \citet{Mat82} described as the
formation of {\em quasar pancakes}.
Towards the
edges of the cloud, the column density is lower, and therefore, there
is a net outward force. 
This can be clearly identified in the
10.93~morb density plot. However, on the very edge, there is a certain
amount of mixing. This will always happen to some extent, as the
contact surface generally has to be resolved by a few grid
cells. Wherever the cloud gas mixes with the ambient gas, the cloud looses
rotational support and falls inwards. There is also some drag due to the 
inflow of the cooling ambient gas.
A Rayleigh-Taylor instability due to the initial cloud acceleration is 
clearly seen in the middle of the cloud.
The cloud continues to expand vertically by the tube of toothpaste
mechanism,  
with lower column density
regions being pushed outwards and  mixing regions coming back inwards.
By 40.08~morb, the cloud has dispersed into three major fragments with
lower density (mixed) filaments falling inwards. 
 The Kelvin-Helmholtz-timescale \citep{Chandra61} is comparable to the evolution time:
\begin{equation}
\tau_\mathrm{KH}=1\,\mathrm{month} 
\frac{\lambda}{\mu\mathrm{pc}}
\left(\frac{\eta}{0.005}\right)^{-1/2} 
\left(\frac{v}{200\,\mathrm{km}\,\mathrm{s}^{-1}}\right)^{-1} \, ,
\end{equation}
where $\lambda$ is the wavelength, $\eta$ the density ratio and $v$
the shear velocity.
As expected, some filaments show the typical rolls of the
Kelvin-Helmholtz instability.
In the final snapshot (54.66~morb), all of the cloud material has
mixed with the ambient gas to some degree, and falls leftwards,
towards the SMBH. Mixing is a resolution effect in our simulations,
but would also happen in reality at scales below our resolution.
The mixing is further illustrated in the density histograms, which we
show for similar snapshot times in Figure~\ref{fig:dhistos}: Between
$t=30.36$~morb and the final snapshot $t=54.66$~morb, the upper
density cutoff drops by about a factor of five, indicating that the
last cloudlet cores have been destroyed by mixing. Consequently, at
this time, we also observe the cloud complex to start falling inwards
as a whole.
During the whole evolution time the entire simulated region stays well
in pressure equilibrium (Figure~\ref{fig:R34o}, bottom left), the
typical deviation being within ten per cent. Also, the magnetic field
in the cloud complex remains close to its initial value of 31~Gauss 
(Figure~\ref{fig:R34o}, bottom middle).
Acceleration is mediated by the compression of the magnetic field. We
expect therefore the velocities to be limited by the Alfv\'en speed,
$c_\mathrm{a}=B/\sqrt{4\pi\rho} \approx 500$~km/s 
in the cloud and $\approx 6000$~km/s in the ambient gas. Indeed, the observed velocities in
the filaments are typically around 30~per~cent of the  Alfv\'en
speed (Figure~\ref{fig:R34o}, bottom right, we use the magnetosonic
speed in the Figures to show that also in the inter-cloud gas, the
velocities remain sub-magnetosonic).

\subsubsection{Underpressured cloud}
We check the dependency on the initial condition in run~R35
(Figure~\ref{fig:R35o}): Here the
field strength is not higher inside the cloud, but set to the same
value as in
the ambient medium at the corresponding radius. The latter is still kept at
$\beta=10$. Also, everything else is as in run~R34. Now, the cloud is
first compressed isotropically due to the initial underpressure before the radial 
(with respect to the black hole)
compression phase. It overshoots and
oscillates in size a few times. The acceleration produces
Rayleigh-Taylor-instabilities on the cloud surface
($t=4.82$~morb). Then the cloud gets compressed anisotropically in
radial (with respect to the black hole) direction
($t=7.23$~morb), and suffers a similar filamentation process than
run~R34, above ($t=9.64$ and~$t=19.29$~morb). 
This initial phase proceeds significantly faster than for R34.
Between $t=19.29$~morb
and $t=37.37$~morb, the remaining dense cloudlet cores are dispersed
(compare Figure~\ref{fig:dhistos}),
and at $t=37.37$~morb, the cloud is infalling towards the SMBH.
The cloud, and respectively its fragments, is quickly compressed to
the equilibrium magnetic field of 31~Gauss, gets quite close to
pressure equilibrium and acquires velocities of about 30~per~cent of
the magnetosonic speed (Figure~\ref{fig:R35o}, bottom plots). Thus,
with the exception of the initial phase, the evolution
is very similar to the one of run~R34.

\subsubsection{Cloud with helical field}
A significantly different result is obtained for a cloud with a
helical magnetic field (R63, Figure~\ref{fig:R63o}). The cloud is
initially not in force equilibrium, and first oscillates in size a few
times. Similar to run~R35, this has caused some filaments by
$t=3.02$~morb. These filaments spread and disperse in much the same
way as for the other runs. However, in contrast to the other runs, the
cloud core remains remarkably stable, and is not significantly
compressed by the radiative, centrifugal and gravitational
forces. This situation remains essentially unchanged until the end of
the simulation ($t=43.55$~morb), which we generally take to be the latest
time before a significant amount of cloud gas has left the grid. 
The final equilibrium is characterised by a total
overpressure of a factor of 1.3 in the cloud core, balanced by the magnetic tension
force (Figure~\ref{fig:R63o}, bottom left). 
In the outer filaments, the magnetic field is of a very
similar magnitude to the other simulations. It is about
ten per cent higher in the cloud core. The velocities are typically about
60~per~cent of the magnetosonic speed, only slightly higher than in the previous
simulations and still sub-magnetosonic.
Also the density histograms (Figure~\ref{fig:dhistos}) are distinctly 
different from the previous runs: Whereas the histograms for the runs without a poloidal
field component did not converge with time, with the upper density cutoff moving
to lower densities at the end of the simulation, the density
histograms for run~R63 stay essentially constant from at least $t=24$~morb up to the
end of the simulation at $t=44$~morb, which corresponds to almost
$1/2$ of the simulation time. The occupied volume is roughly given by
\eq{\frac{\mathrm{d}V}{\mathrm{d\,log}\,\rho} \propto \rho^{-1}\, ,}
and also extends to densities exceeding the initial cloud density by a
factor of ten.

\begin{figure*}
\centering
\includegraphics[width=\textwidth]{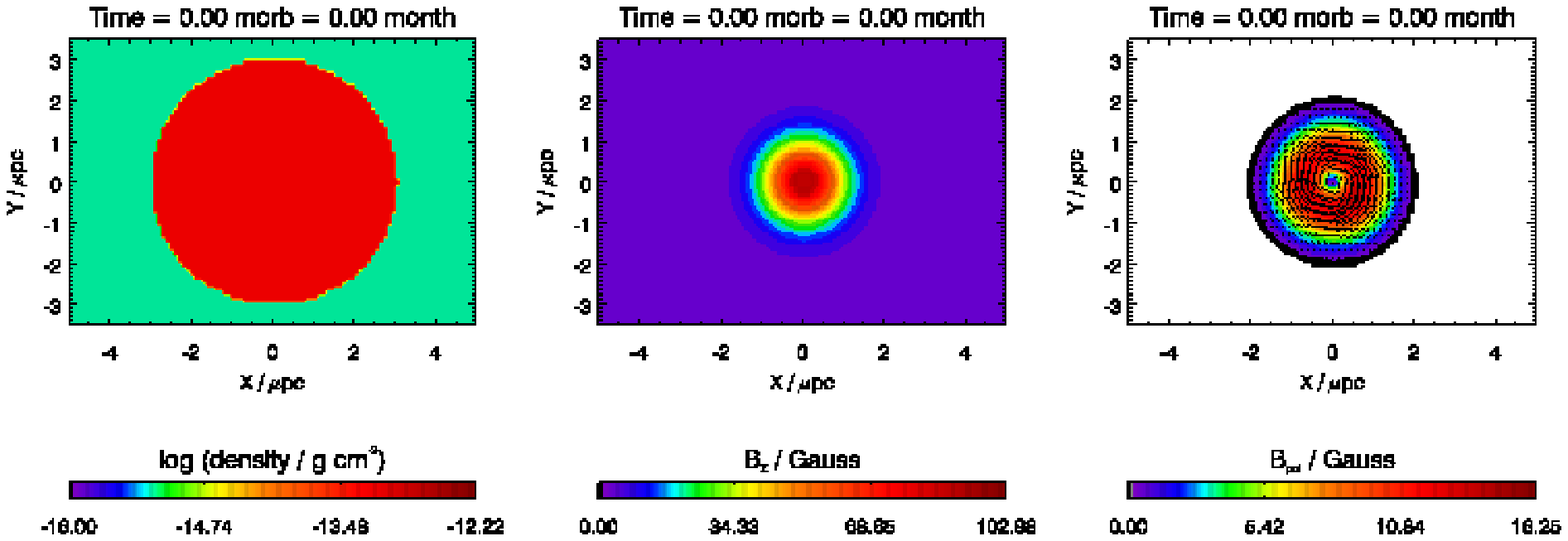}
\includegraphics[width=\textwidth]{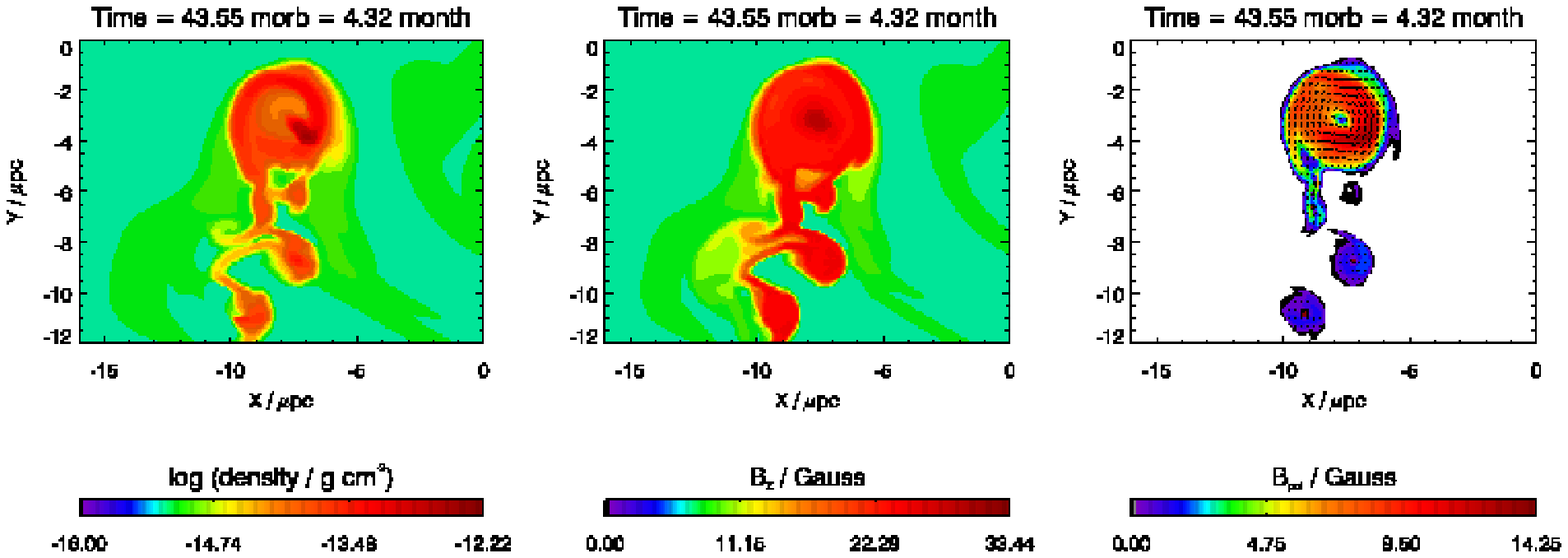}
\caption{\small Close-up of the cloud core of run~R63 at the beginning
  (top row) and at the end (bottom row) of
  the simulation. The logarithm of the density is shown on the left,
  the azimuthal magnetic field in the middle and the magnitude of the
  poloidal field with overlaid poloidal field vectors on the
  right. The coordinates are relative to the initial cloud centre. The
slight upwards shift at the end of the simulation is due to an
interaction with the boundary, which defines the end of the simulation.}
\label{fig:R63close}
\end{figure*}

Close-ups of the cloud core at the beginning and at the end of the simulation 
are shown in Figure~\ref{fig:R63close}. The cloud has essentially the
same size as in the beginning. The density is however no longer
homogeneous, but displays a round filamentary pattern, following the
cloud surface. 
The magnetic field geometry is essentially identical to the one of the
initial condition: a dominantly azimuthal core with a helix winding
around. The peak value of the azimuthal field has however reduced by a
factor of three, whereas the poloidal field strength remained almost unchanged. 

Interestingly, many of the filaments that the cloud sheds are essentially
devoid of poloidal field. Due to the solenoidal condition, filaments
shed by the cloud need to have a bidirectional magnetic field. We
have observed this at 0.1~G level near the cloud surface. But
since they are often at the resolution limit, the poloidal field
tends to cancel numerically, while the toroidal one is unaffected by this.
Without poloidal field in the filaments, they disperse much like the
filaments in the other simulations. 
Yet, sometimes filaments are shed which are thick enough to keep their
poloidal field. Two such examples are present in
Figure~\ref{fig:R63close}. Interestingly, they also adopt the field
geometry of the parent cloud. This indicates that we may have found a
magnetic field configuration for such clouds.
It is clear that a poloidal field stabilises the cloud significantly.

%%%%%%%%%%%%%%%%%%%%%%%%%%%%%%%%%%%%%%%%%%%%%%%%%%%%
%%%%%%%%%%%%%%%%%%%%%%%%%%%%%%%%%%%%%%%%%%%%%%%%%%%%
%%%%%%%%%%%%%%%%%%%%%%%%%%%%%%%%%%%%%%%%%%%%%%%%%%%%
\begin{figure*}
\centering
\includegraphics[width=\textwidth]{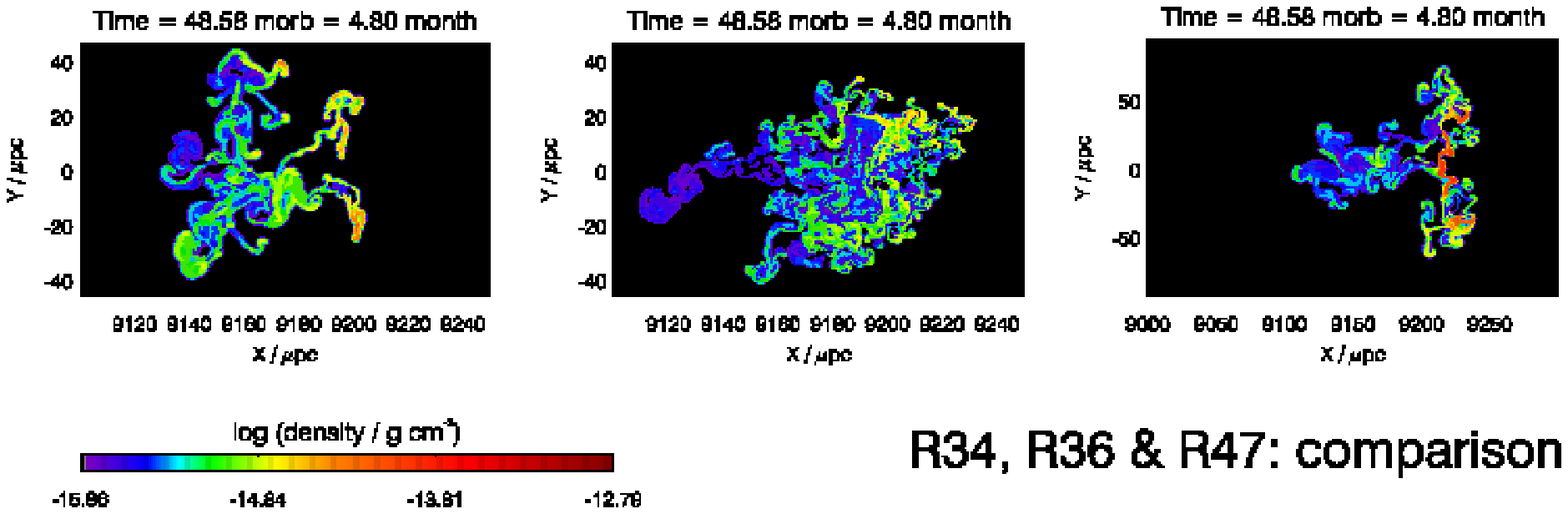}
\caption{\small  Comparison of the density distribution of run~R34
  (left) to run~R36 (double resolution, middle) and run~R47 (double
  spatial scales, right) at the common snapshot time of 4.58~morb.}
\label{fig:rescomp}
\end{figure*}
%%%%%%%%%%%%%%%%%%%%%%%%%%%%%%%%%%%%%%%%%%%%%%%%%%%%
%%%%%%%%%%%%%%%%%%%%%%%%%%%%%%%%%%%%%%%%%%%%%%%%%%%%
\begin{figure*}
\centering
\includegraphics[width=\textwidth]{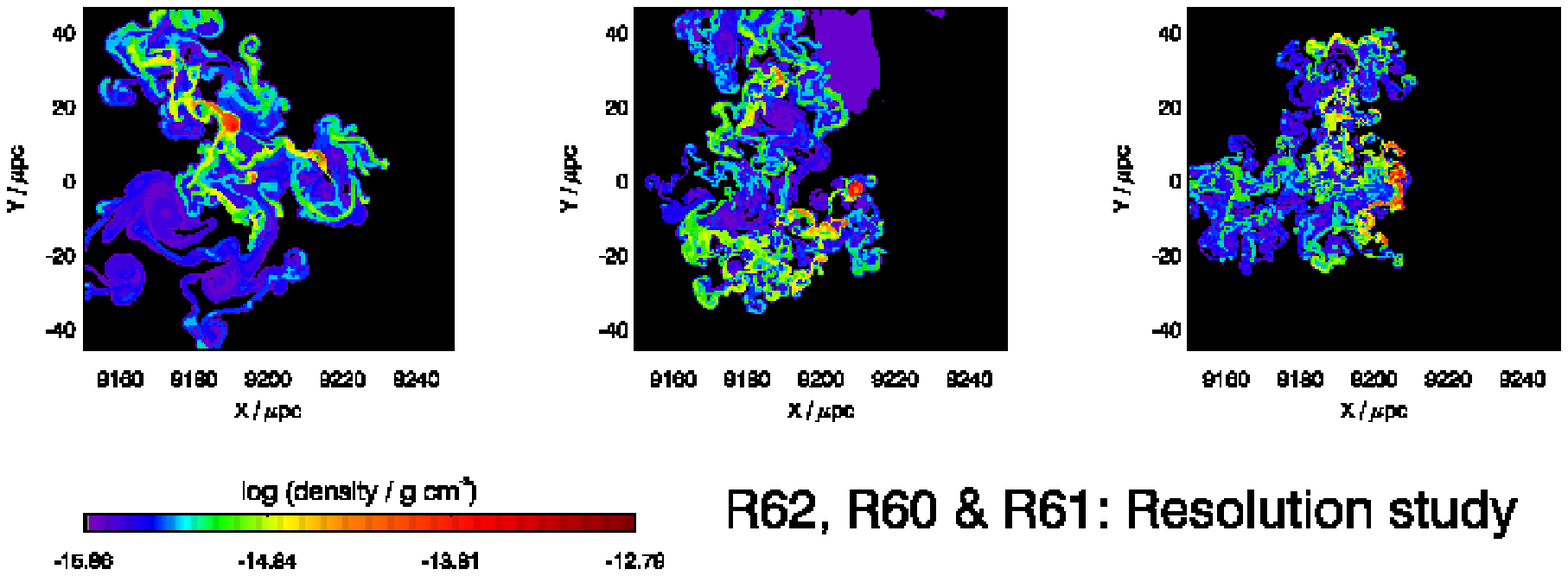}
\caption{\small  Comparison of the density distribution of run~R60
  (middle, smaller grid but otherwise identical to R63) to run~R62 
  (half resolution, left) and run~R61 (50 per cent higher resolution, right) at the common snapshot time of 48.58~morb.}
\label{fig:rescomppol}
\end{figure*}
%%%%%%%%%%%%%%%%%%%%%%%%%%%%%%%%%%%%%%%%%%%%%%%%%%%%

\subsection{Resolution dependence}
 We have re-simulated R34 at twice the spatial resolution (R36,
Figure~\ref{fig:rescomp}, middle). The evolution is quite similar to run~R34,
with the exception that the dispersion of the cloud and cloudlets
happen faster. This may be seen from the more uniform appearance of
the density distributions (Figure~\ref{fig:rescomp}), but also from
the density histograms (Figure~\ref{fig:dhistos}), which are shown at
the same timesteps: While the histograms at the corresponding times
are generally very similar, the density cutoffs at $t=54.66$~morb
differ by about 0.2~dex. Thus, the simulation can be regarded as
overall numerically converged, but there is increased turbulent mixing at small
scales for increasing resolution.

\begin{figure}
\centering
\includegraphics[width=.49\textwidth]{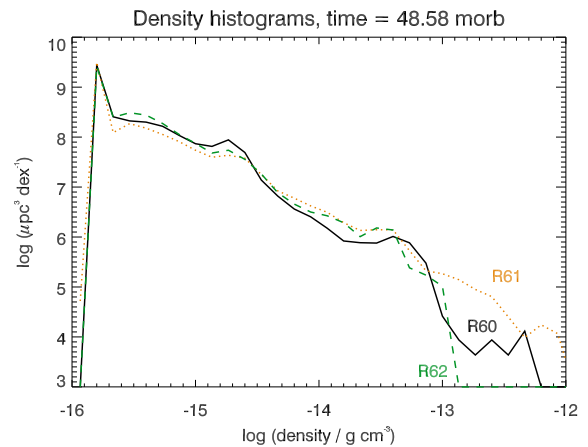}
\caption{\small  Density histograms for runs~R60
  (solid black) R61 (dotted yellow,  50 per cent higher resolution)
  and R62 (dashed green, half resolution) at 48.58 morb. The high density part
  disappears at low resolution.}
\label{fig:rescomppoldh}
\end{figure}

\change{We have also studied the resolution dependence for run R63. For this
study, we have used a smaller grid, because we have found that reflection of sound
waves at the grid boundaries somewhat influences the motion of the
filaments. R60 is the re-simulation of R63 at the same resolution. }
R62 uses 50 per cent less cells on a side, R61 has 50 per
cent more cells on a side. Logarithmic density plots of all three runs
at $t=48.58$~morb are shown in Figure~\ref{fig:rescomppol}.
The filamentary systems are in general quite similar, though they have
of course a finer structure at higher resolution. Yet, their extent
and therefore kinematics is similar. The cloud core \change{has} 
drifted outward slightly for the two higher
resolution runs. It is less dense and moving inwards for the low
resolution run. The resolution is in this case not enough to resolve
the cloud core structure, and therefore, the clouds disperses
faster. This is also evident from the density histograms
(Figure~\ref{fig:rescomppoldh}). The low resolution run lacks cloud
material at high densities, which the two higher resolution
clouds do show. It is interesting that this is in a way opposite to
the case without poloidal field. There, higher resolution had made the
cloud disperse quicker. Here, high resolution is essential to preserve
the cloud core at high density. Our results are thus robust under
resolution changes.

\subsection{Scaling with cloud size}
The simulations are not scalable, as the source terms introduce
a dimensional scale to the problem.
In order to explore the scaling behaviour, we re-simulated run~R34 with
all scales doubled (R47, Figure~\ref{fig:rescomp}, right). As
expected, the evolution now takes longer, but not nearly twice as long
as for run~R34. Otherwise, we find a very similar behaviour: Again the
cloud is first compressed radially, then expands vertically by the tube
of toothpaste mechanism, disperses into a filamentary system and then
moves inwards due to the mixing related loss of rotational support and
drag by the cooling infalling ambient gas.

\subsection{Turbulent velocities}
We define the mass-weighted root mean square poloidal velocity (rms
velocity) by 
\eqs{v_\mathrm{rms}^2 &=&
  \frac{\sum_m\mathrm{d}m\, ({\bf v}_\mathrm{p}-\bar{\bf
        v}_\mathrm{p})^2} {\sum_m \mathrm{d}m}\\
    \bar{\bf v}_\mathrm{p}&=&
\frac{\sum_m\mathrm{d}m\, {\bf
    v}_\mathrm{p}}{\sum_m\mathrm{d}m}\, ,}
with the poloidal velocity vector ${\bf v}_\mathrm{p}$.

\begin{figure}
\centering
\includegraphics[width=.5\textwidth]{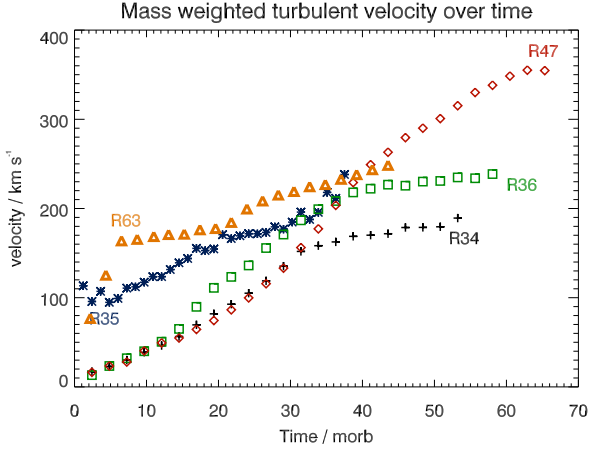}
\caption{\small Mass-weighted root mean square velocities for runs
R34 (black plus signs), R35 (blue stars), R36 (green squares), R47
(red diamonds) and R63 (orange triangles)
over time. Only the poloidal velocity component is taken into account.}
\label{fig:vrms}
\end{figure}

The dependency of $v_\mathrm{rms}$ 
on time is shown for all runs in Figure~\ref{fig:vrms}. The clouds set up in
pressure equilibrium show first a shallow rise and then stay roughly
constant. From the density histograms (Figure~\ref{fig:dhistos}), the
acceleration decreases when the densest cores are completely shredded.
Hence, it seems that kinetic energy may be generated as long as dense
cloud material may be squeezed radially and pushed out
vertically. 
Since run~R36 has a higher resolution than run R34, the gas may be
compressed further. More energy is therefore stored temporarily in the
magnetic field, which ultimately results in higher velocities.
Run~R47 has twice the spatial scales but is otherwise identical to
run~R34. The net force per unit volume has not changed. Since the
initial densities are also the same, the accelerations in each cell are the
same, too. Therefore, the turbulent velocities increase in a very similar
way up to about 35~morb. Since the velocities are the same, but the
length scale is twice as long in run~R47, there is still room for more
compression in simulation~R47 at this time. Consequently, the
turbulent velocity increases further, until the resolution limit is
reached near 57~morb. One may understand the final turbulent
velocity in terms of the total available space for compression:
The gravitational and centrifugal forces act at similar strength in
every cloud cell. The radiative force is concentrated on the
illuminated surface. The net compression force is roughly $F= (1-V^2)
F_\mathrm{G}$, where $V$ is the initial rotation velocity of the cloud
in Kepler units.
The total energy transfered to the cloud in the limit of complete
linear compression would therefore be:
\eq{E_\mathrm{comp}=\frac{1-V^2}{2}\pi F_\mathrm{G} R_\mathrm{cld}\, ,}
where the numerical factor is due to the average cloud thickness of
$\pi R_\mathrm{cld}/2$.
If this energy would be completely used for accelerating the cloud, a
turbulent velocity of:
\eqs{v_\mathrm{t,c} &=& 
  \left( \pi (1-V^2) R_\mathrm{cld} \frac{G M_\mathrm{BH}}{r_\mathrm{cld}^2}\right)^{1/2}\\
&=& 115\, \mathrm{km\, s}^{-1} \nonumber
\left(\frac{M_\mathrm{BH} }{10^8 M_\odot}\right)^{1/2}
\left(\frac{r_\mathrm{cld} }{9200\,\mu\mathrm{pc}}\right)^{-1}
\left(\frac{R_\mathrm{cld} }{3\,\mu\mathrm{pc}}\right)^{1/2}}
would result for $V=0.85$.
As may be seen from Figure~\ref{fig:vrms}, the velocities are somewhat
higher than this estimate. This is likely due to the increased
radiative force during the lateral expansion of the clouds.

The clouds not set up in pressure equilibrium (R35 and R63) suffer initially a very
quick increase of the rms velocity, as might have been expected. They
evolve quickly towards pressure equilibrium. Then, as in the initial 
pressure equilibrium runs, the turbulent velocities change only gradually. 
Several curves show an upturn towards the end, which is due to the
beginning interaction with the grid boundary. 

%%%%%%%%%%%%%%%%%%%%%%%%%%%%%%%%%%%%%%%%%%%%%%%%%%%%
%%%%%%%%%%%%%%%%%%%%%%%%%%%%%%%%%%%%%%%%%%%%%%%%%%%%

%%%%%%%%%%%%%%%%%%%%%%%%%%%%%%%%%%%%%%%%%%%%%%%%%%%%
%%%%%%%%%%%%%%%%%%%%%%%%%%%%%%%%%%%%%%%%%%%%%%%%%%%%

\section{Discussion}\label{disc}
\change{We have investigated whether magnetic fields might
  contribute to the stability of BLR clouds. Stability is an issue in
  models without magnetic fields (compare section~\ref{intro}). 
  We first combine literature data to
  get an idea of the magnitude of magnetic and other forces 
  that might be relevant in this
  context. 
  We find that reverberation mapping data \rvtwo{is consistent with} the bound cloud
  picture for a reasonable column density. Yet, the forces of radiation and
  gravity are comparable. We argue that the magnetic energy density,
  which sets the magnitude of the magnetic pressure and tension forces,
  should be at least of the same order of magnitude as the radiation pressure,
  in order to be at all able to contribute to the stability of the
  cloud. Estimating these parameters from the literature 
  turned out to be complicated by the fact that the flux of
radiation inferred via reverberation mapping studies \rvtwo{leads to a
rather high photoionisation parameter.} One could take this as indication that
there are no clouds at all and cloudless models like the one of
\citet{Murea95} should be preferred. 
\rvtwo{Yet, for cloud densities on the high side of the allowed
  values, which would also mean that the typical BLRS would more likely be
  gravitationally bound, or when adopting other changes of the
  relevant parameters within the range allowed by the uncertainties,
  the photoionisation parameter may be brought down to a reasonable value.}

%We have however pointed out that
%the upper bound for the ionisation parameter $\Xi$ is not very well
%constrained, and that an increase of $\Xi$ as well as the cloud
%density as suggested by \citet{Ferlea92} might be sufficient to
%explain the data. 

\rvtwo{A gravitationally bound cloud model} entails the problem that the radiation leads
to a surface force whereas gravity and centrifugal forces are body
forces. Hence, an equilibrium configuration with these forces alone
may not be found for the aforementioned assumptions.
As all these forces act only radially away or towards the
SMBH, an isotropic thermal pressure that was to stabilise the cloud
against the compressive forces would necessarily also 
cause lateral expansion. Magnetic
forces are anisotropic and are therefore in principal suited to
counteract the compression and stabilise the clouds. We have performed
MHD simulations with different magnetic field configurations to test
this idea.}

With an azimuthal magnetic field, only, 
the dynamics is dominated by the open tube of toothpaste
mechanism: radial compression is followed by vertical
expansion. Kelvin-Helmholtz and column density instabilities together
with repeated, but smaller, tube of toothpaste squeezing events transform the
initially homogeneous cloud quickly into a filamentary system. 
This result is in good agreement with the unstable pancake picture
\citep{Mat82,Mat86}. In fact, in this case the effect of the magnetic
field is just an additional pressure. Hence, the result remains
essentially the
same as in the pure hydrodynamic case.
The
cloud fragments are shredded down to the resolution level. Mixing with
the pressure supported ambient gas reduces the rotational support, and
the clouds fall inwards. Additionally, the atmosphere is not entirely
stable and features some inwards flow due to the cooling.
The timescale for this process is given first
by
the cloud compression timescale:
\eqs{t_\mathrm{c}&=&\left(\frac{4R_\mathrm{cld}}{(1-V^2)g/2}\right)^{1/2}\\
 &=& 4 \,\mathrm{months}\,
\left(\frac{M_\mathrm{BH} }{10^8 M_\odot}\right)^{-1/2}
\left(\frac{r_\mathrm{cld} }{9200\,\mu\mathrm{pc}}\right)
\left(\frac{R_\mathrm{cld}}{3\mu\mathrm{pc}}\right)^{1/2}\, ,\nonumber}
where $g$ is the gravitational acceleration, and the evaluation is again for an initial
rotation velocity $V=0.85$.
Our numerical simulations show that clouds with an azimuthal magnetic
field, only, are dispersed to the resolution level, and hence rapidly
dissolve in the ambient medium within about $t_\mathrm{c}$, which is
also similar to the Kelvin-Helmholtz timescale for the simulated cases.
We show that the dispersal is even faster for higher resolution.
The mixing, on the other hand, depends on the ambient density, via
the Kelvin-Helmholtz timescale. It is therefore expected to be slower
for lower ambient densities. Not much is known about the density in
the inter-cloud gas. Assuming that the density would be lower by a
factor of $10^4$, would increase the Kelvin Helmholtz timescale by a
factor of 100, to the range of the orbital timescale. This would lead
to extremely spread out pancake clouds, similar to the ones envisaged
by \citet{Mat86}. We have also tried a factor of ten lower inter-cloud
density than reported here. In this case, the cloud spreads 
quite quickly over the grid boundary probably due to the
diminished resistance of the ambient gas, whereas the density
histograms showed more dense gas at comparable times, as expected.
In three dimensions, such pancake clouds would also face the ram pressure,
which might also lead to compressed cloud heads with cometary tails,
and a stability timescale of order the orbital timescale \citep{Mat86}.

The runs with azimuthal magnetic field, only, underline the stability
problem, and demonstrate clearly that our simulations are able to
capture the destruction process adequately.
Our runs with a helical magnetic field result in a much more stable cloud
configuration, which survives the destruction process for the
simulation time. The cloud corresponds to a helical filament in
azimuthal direction. The core of this filament is dominated by the
azimuthal field, producing an overpressure of about
30 per cent. This is balanced by the magnetic tension force of the poloidal
field surrounding the core. The stability of the configuration results
from the nature of the tension force which is inversely proportional
to the curvature radius of the field lines. A radial compression
straightens the field lines vertically (less radial tension force) and
bends them more sharply at the vertical ends (increased vertical
restoring tension force). A similar restoring force is obtained for
compressions in every other direction. This configuration is very similar to the
magnetic field used in tokamaks. One reason why it is employed in this
context is precisely its MHD stability. 

\change{In order to isolate the effects of the magnetic field, we have
used a negligible thermal cloud pressure.}
In our simulations we have \change{realised this via} a minimum temperature of
1000~K. The simulations would not have been any different had we used
$10^4$ or even $10^5$~K, as the magnetic pressure would still be by
far dominant in the clouds. We had initially decided for a low minimum
temperature to check if the induced dynamics would lead to any
significant increase of the cloud temperature \rvtwo{via shock
  heating. The balance between shock heating and radiative cooling
  was found by \citet{KA07} in their simulations of clouds in jet
  cocoons to produce a peak in the temperature histograms around
  $10^4$~K. We do not find this effect here:} 
Even for the late phases of R34, where the cloud is already
essentially dispersed, the by far major part of the filaments 
remain close to the minimum temperature. Therefore, we do not find a
heating mechanism which could compete with photoionisation.

\change{Cloud stability in the limit of negligible magnetic fields has
  been studied extensively in the past (compare
  section~\ref{intro}). We have studied here the case of a dominant
  magnetic field compared to the thermal energy density, and separated
  the effects of magnetic pressure and magnetic tension. 
  The azimuthal field behaves like a pressure in our axisymmetric
  simulations. Hence, we expect that we could in principle easily accommodate for a
  thermal pressure comparable to the magnetic one by replacing some or
  all of the magnetic pressure due to the azimuthal field by a
  corresponding thermal pressure. Details would of course depend on
  the effective equation of state, as the thermal pressure is affected
  by heating and cooling processes.}

Another question one has to ask is how likely it would be to find such
a helical configuration around real AGN. If the inter-cloud density
would be comparable to the one assumed in our simulations, 
all unstable configurations would be shredded
and dissolved within months. It is then quite difficult to replace the
clouds quickly enough. If they would for example be launched from the
accretion disc by a wind with velocities comparable to ones observed
in the BLR, it would take a similar amount of time to be lifted to a
significant altitude above the disc. Yet, the BLR cannot be a thin
disc \citep[e.g.][]{Ost78}. Our helical field simulation has started
quite far away from the final equilibrium, yet it has reached a stable
configuration. It seems therefore possible
that many different kinds of initial conditions eventually find an
equilibrium solution. It is beyond the scope of this study how many
stable filaments may be produced in given conditions of the
interstellar medium. 
% It is however clear that the radiation of the AGN
% does not make things worse: If a given filament would be dispersed by
% the radiative mechanism in the vicinity of an AGN, it would have been
% destroyed only slightly later by the tidal mechanism without the
% radiation.
% Therefore, if gas filaments are present in the vicinity of a SMBH,
% they are likely to survive also when the AGN lights up.

\rvtwo{Regarding the absolute magnitude of the magnetic field,
  \citet{Rees87} has already pointed out that magnetic field 
  values of a few Gauss, as required here, are expected in BLRs due to relativistic
  winds, accretion flows or accretion driven winds. He also mentions
  the possibility of magnetic fields within the clouds if they are
  created by the thermal instability. We note that a similar value for the magnetic
field has recently been estimated from their possible effects on the
polarisation of the H$\alpha$ line \citep{Silea12}.}

What level of internal turbulence do our simulations predict?
We have argued that the available energy for turbulent motions is
limited by the cloud compression and the Alfv\'en speed. Since the initial cloud volumes are
unknown, we cannot make a quantitative prediction from this argument.
If the BLR cloud population would be dominated by unstable clouds,
being constantly shredded and replaced, we would predict turbulent
velocities of about a third of the magnetosonic speed. If the
radiation pressure would indeed be matched by the magnetic pressure in
the cloud, the magnetosonic speed would be of order 1000~km/s, and the
predicted internal velocities consequently a few hundred. For stable
clouds of the helical type as shown above, we would expect
even less in the cloud cores. Hence, the turbulent
velocities of about 1000~km/s derived by \citet{KolZet11} cannot be
reached in this way.

The cometary clouds observed by \change{\citet{Maiea10}}, with dense cores and
filamentary tails are compatible with our simulation
results. Taking the orbital motion into account, the filamentary tails
would be dominantly elongated in the azimuthal direction, as probably
also in the
observations. If these clouds would indeed be destroyed within a few
months, as estimated by \citet{Maiea10}, this would fit exactly with
the destruction timescale of the radiative destruction mechanism
discussed here. 

\change{In the literature one may find also BLR models that involve no clouds
at all, e.g. the disc wind model of \citet{Murea95}. Here, the BLR
forms the base of a wind accelerated by radiative and thermal
pressure. The advantages and disadvantages of models with and without
clouds have been nicely summarised by \citet{Net08}: Cloud models
provide a significantly better match to the emission line structures,
while cloudless models obviously avoid the confinement problem altogether.}

We have addressed the meridional stability problem, only. While this
is certainly one of the major issues in the BLR cloud stability
problem, we have not shown that the cloud would also survive 3D
effects. From the MHD point of view, kink instabilities are likely to
occur, which would cause bends in the cloud filament along its major
axis. 
\rvtwo{Further, the clouds are unlikely to be extended along the entire
azimuthal angle as necessarily implied in our axisymmetric
simulations. In reality, the clouds could still be elongated
substantially in the azimuthal direction, if the ends would be
anchored in a thinner disc structure, similar to coronal loops on the
surface of the sun. In simple magnetic field configurations with
closed field lines within the cloud, the field could of course be
transformed via reconnection, which would likely lead to cloud
splitting as each closed field region could support a cloud
on its own. An azimuthally extended cloud with the azimuthal field
closing at the ends and returning through some part of the cloud would
be expected to expand in the polar direction and form something like a
closed ring. 
 For any non-axisymmetric clouds,
also} the ram pressure would be significant. For our cloud
setups the azimuthal ram pressure would be comparable to the magnetic pressure
in the cloud. If one can find magnetic field configurations that would
also stabilise against this effect is beyond the scope of the present
investigation.

In order to reach pressure equilibrium with the clouds, also the
inter-cloud medium should be magnetised on a similar level. This might
be accessible to Faraday rotation. The expected rotation measure would
be of order
\eq{\mathrm{RM} = 8.12\times 10^8 \int_0^L n_\mathrm{e,6}  {  B} \cdot
  \mathrm{d}l \; \mathrm{rad\, m^{-2}},}
where the path-length d$l$ is measured in units of $10^{-3}$~pc, $B$ in Gauss and
the electron density in units of $10^6$~cm$^{-3}$. As mentioned above,
the density is uncertain, and could also be a few orders of magnitude
lower.
For a turbulent
medium, one would therefore predict that signals below $\approx
10^{13}$~Hz ($\lambda>30\mu$m) would be depolarised \citep*{KAB2007}. 
This prediction
might therefore be tested in the future by infrared polarimetry, if
one is sure to observe emission from a jet base.

\change{
For our simulations, a change in the geometry of the magnetic field 
in the inter-cloud region
should not matter much, because the thermal energy density 
dominates there over the magnetic one. As long as the forces are set up close to
equilibrium, we would expect an outcome in between runs R34 (initial
pressure equilibrium) and R35
(no initial pressure equilibrium). In reality, the situation might of
course be more complex, and the inter-cloud medium might be
magnetically dominated and at the same time \rvtwo{have} a non-zero poloidal
component. It would then be important if the magnetic field would be
topologically connected to the clouds. In that case, the cloud gas
could escape along the field lines. It might also happen that regions
of space with a favourable magnetic field configuration might protect
new clouds which are just forming from the thermal instability. Yet,
these complexities are beyond the scope of the present discussion.

Summarising, we find some indication that bound BLR clouds might
indeed be stabilised by the magnetic field  against collapse due to 
opposing radiative and
gravitational forces. In this
case, the magnetic field would have to have a poloidal component.
}

\section{Conclusions}\label{conc}

\change{
Gravitationally bound clouds facing strong radiation
pressure are unstable in the purely hydrodynamic case because
radiative and gravitational forces compress the clouds radially,
whereas the thermal pressure acts isotropic. They may be more stable if they are significantly
magnetised. In particular, we find in axisymmetric magnetohydrodynamic
simulations with a prescription for the radiation pressure that 
the magnetic tension force
produces more stable clouds, while in a situation where the geometry of
the magnetic field is such that its effect is only analogous to an
additional pressure, the clouds are similarly unstable as in the
hydrodynamic case. In order to be effective, in the BLRs of a
discussed sample of AGN, the magnetic field strength should be 
\rvtwo{of
order a few Gauss, accurate to about an order of magnitude and} 
constrained by the condition that magnetic,
radiative and gravitational forces should be comparable.}

\section*{Acknowledgements}
We thank the anonymous referee whose comments contributed
significantly to improve the manuscript. This work has been supported
by an MPG fellowship.

\bsp
\bibliographystyle{mn2e}
\bibliography{/Users/mkrause/texinput/references}
%\bibliography{/Users/mghk/Desktop/bttry}

\label{lastpage}

\end{document}